\documentclass[final,5p,times,twocolumn,authoryear]{elsarticle}

\usepackage[T1]{fontenc}
\makeatletter\@ifclasswith{elsarticle}{times}{}{\usepackage{lmodern}}\makeatother  
\usepackage{amsmath,amssymb}
\usepackage{graphicx}
\usepackage{bm}
\usepackage{booktabs}
\usepackage{array}
\usepackage{adjustbox}
\usepackage{xcolor}
\usepackage[colorlinks=true,breaklinks=true,linkcolor=blue!60!black,citecolor=blue!60!black,urlcolor=blue!60!black]{hyperref}

\usepackage{etoolbox}
\AtBeginEnvironment{table}{\setlength{\belowcaptionskip}{6pt}}
\AtBeginEnvironment{table*}{\setlength{\belowcaptionskip}{6pt}}
\newcommand{\LCDM}{$\Lambda$CDM}
\newcommand{\Om}{\Omega_{m,0}}
\newcommand{\Or}{\Omega_{r,0}}
\newcommand{\dchi}{\Delta\chi^2}

\newlength{\colfigwidth}
\AtBeginDocument{\ifdim\columnwidth<0.6\textwidth\setlength{\colfigwidth}{\columnwidth}%
  \else\setlength{\colfigwidth}{0.72\textwidth}\fi}

\journal{High Energy Astrophysics}

\begin{document}

\begin{frontmatter}

\title{Phantom crossing and cosmological constraints in a Hu--Sawicki-inspired $f(Q)$ model}

\author[sapienza]{Giovanni Tuccio\corref{cor1}}
\ead{tuccio.1917847@studenti.uniroma1.it}
\cortext[cor1]{Corresponding author}

\author[sapienza,enea]{Giovanni Montani}
\author[sapienza,infn]{Alessandro Melchiorri}

\affiliation[sapienza]{organization={Physics Department, Sapienza University of Rome},
  addressline={P.le A.Moro 5},
  postcode={00185},
  city={Roma},
  country={Italy}}

\affiliation[enea]{organization={ENEA,Nuclear Department, C.R. Frascati},
  addressline={Via E. Fermi 45},
  postcode={00044},
  city={Frascati (Roma)},
  country={Italy}}

\affiliation[infn]{organization={Istituto Nazionale di Fisica Nucleare (INFN), Sezione di Roma},
  addressline={Piazzale Aldo Moro 5},
  postcode={00185},
  city={Roma},
  country={Italy}}

\begin{abstract}
We propose an extension of nonlinear symmetric teleparallel gravity inspired by the Hu–Sawicki
model, $f(Q)=Q+2\Lambda+\alpha\mu^2(x-1)/(1+x)^2$, with $x=Q/\mu^2$ where a single dimensionless parameter
$\alpha$ controls the departure from \LCDM. The effective dark-energy equation of state crosses the
phantom divide at $x_\pm=3\pm2\sqrt2$ whatever the value of $\alpha$, as happens for any deformation
linear in $\alpha$, while the redshift of the crossing is fixed by the scale $\eta=\mu/H_0$. For $\alpha$ > 0 the
crossing is from phantom to quintessence-like behaviour, in the same direction as the DESI fits.
Asking for a regular background for every $Q$ ($f_Q>0$ and an invertible Friedmann constraint)
limits the deformation to $-1/3<\alpha<0.962$. We test the background expansion against cosmic
chronometers, DESI DR2 baryon acoustic oscillations, Planck distance priors and supernovae
(Pantheon+, DES-Y5 with the Dovekie recalibration, and Union3), and include $f\sigma_8$ data only as
an exploratory extension.
For the reference scale
$\eta=\sqrt6$, fixed a priori, the best fit has $\alpha>0$ for all three supernova samples ($\alpha=0.085\pm0.060$
with Pantheon+) and $\dchi=2.0$--$2.9$ with respect to \LCDM\ ($1.4$--$1.7\sigma$); the Bayes factor and the Bayesian
information criterion mildly favour \LCDM, so the data remain compatible with $\alpha=0$. If the scale is chosen by
the data, the improvement grows to $\dchi=4.8$--$6.9$ at $\eta=1.6$; Monte Carlo simulations under \LCDM\ that repeat
the search over $\eta$ give a significance of $2.1$--$2.5\sigma$. Among the one-parameter extensions considered
($w$CDM, power-law and exponential $f(Q)$) the model gives the largest improvement, between a sixth and a quarter of
that of the two-parameter $w_0w_a$ description. The analysis does not address the strong coupling of $f(Q)$
perturbations, and lunar laser ranging requires a screening mechanism not yet known for $f(Q)$.
\end{abstract}

\begin{keyword}
Modified gravity \sep $f(Q)$ gravity \sep Dark energy \sep Phantom crossing \sep Cosmological constraints
\end{keyword}

\end{frontmatter}

\section{Introduction}
\label{sec:intro}

General relativity is a remarkably successful theory, although the observed late-time acceleration of the
Universe~\citep{Riess1998,Perlmutter1999}  motivate extensions in which the dark sector may have a dynamical
origin. Interpreting this acceleration as a cosmological constant is phenomenologically successful but raises the
cosmological constant problem~\citep{Weinberg1989}. Dynamical dark energy, including equations of state below the
phantom divide $w=-1$~\citep{Melchiorri2003}, has been constrained with increasing precision~\citep{DiValentino2017DE,
DiValentino2020Planck}, and interactions in the dark sector have been proposed to address the Hubble
tension~\citep{DiValentino2017IDE,DiValentino2020IDE,DiValentino2021}. Over the last decades several modified-gravity
frameworks have been developed, including curvature-based $f(R)$ theories~\citep{Nojiri2017,DeFelice2010}. In the late
Universe, metric $f(R)$ gravity has been proposed as a dynamical origin of the decrease of the effective Hubble constant
with redshift found in binned analyses of type Ia supernovae~\citep{Dainotti2021,Dainotti2022,Dainotti2026}, both in the
Jordan frame~\citep{Schiavone2023,Schiavone2025,Valletta2026} and in the presence of an evolving dark-energy
component~\citep{Montani2023MNRAS,Montani2025decay}. The redshift trend of the effective Hubble constant has also been proposed as a marker
of the nature of dark energy, an increasing trend indicating a quintessence-like and a
decreasing one a phantom-like behaviour~\citep{Fazzari2026}. Other frameworks include torsion-based $f(T)$
models~\citep{Cai2016fT} and nonlinear extensions of the symmetric teleparallel equivalent of general relativity
(STEGR)~\citep{BeltranJimenez2018CGR,Heisenberg2024,BeltranJimenez2020Cosmo}. The latter is built on a quantity called non-metricity,  that describes how the length of vectors changes under parallel transport. STEGR is equivalent
to general relativity, but its nonlinear extensions $f(Q)$ lead to phenomenology that is not captured by $f(R)$
models. Several $f(Q)$ models have been confronted with observations~\citep{Anagnostopoulos2021,Lazkoz2019,
Barros2020}, generally finding that current data allow for small deviations from \LCDM.

A useful technical feature of $f(Q)$ cosmology is that in the coincident gauge with Cartesian coordinates the flat
Friedmann--Lema\^itre--Robertson--Walker (FLRW) background is described entirely by $Q=6H^2$. This is not the only
possibility: three connection branches are compatible with a spatially flat FLRW metric~\citep{Dimakis2022}
\citet{Jensko2025} has shown, starting from the cosmological Killing vectors, that all of them can be described in the
coincident gauge with suitably adapted coordinates. It has recently been shown that perturbations around
$f(Q)$ cosmologies are problematic in all of them~\citep{Gomes2024}. We return to this point in Sec.~\ref{sec:discussion}: our main results concern the
background expansion; the quasi-static growth of matter perturbations is considered only as an exploratory extension
(Sec.~\ref{sec:growth}), conditional on a consistent completion of the theory.

Within this framework, we consider a one-parameter extension of \LCDM\ inspired by the Hu--Sawicki model of $f(R)$
gravity~\citep{HuSawicki2007},
\begin{equation}
f(Q)=Q+2\Lambda+\alpha\mu^2\,\frac{x-1}{(1+x)^2}\,,\qquad x\equiv\frac{Q}{\mu^2}=\frac{6H^2}{\mu^2}\,,
\label{eq:model}
\end{equation}
with one dimensionless deformation parameter $\alpha$ and one characteristic scale $\mu$. We adopt the sign convention
in which general relativity corresponds to $f(Q)=Q$ and \LCDM\ to $f(Q)=Q+2\Lambda$~\citep{BeltranJimenez2020Cosmo}.

In the coincident gauge the background equations of $f(Q)$ gravity coincide with those of $f(T)$
gravity~\citep{BeltranJimenez2020Cosmo,Yang2025quintomDR2}, so background results carry over between the two. Crossings of the phantom divide
have been found in viable $f(R)$ models~\citep{AmendolaTsujikawa2008}, in $f(T)$~\citep{WuYu2011,Bamba2011fT} and in
$f(Q)$~\citep{AroraSahoo2022}, and quintom behaviour in $f(T)$ and $f(Q)$ has been reconstructed from DESI
data~\citep{Yang2024quintom,Yang2025quintomDR2}; other $f(Q)$ models have been confronted with DESI
DR2~\citep{Paliathanasis2025,Mazumdar2025}. What is specific to the model studied here is the form of the deformation,
which places a single crossing on the past trajectory for $1.01<\eta<5.9$, in the direction preferred by the DESI
$w_0w_a$ fits: phantom in the past and quintessence-like today for $\alpha>0$.

This work focuses on three points.
\begin{itemize}
\item 
In this formulation,$\rho_{\rm DE}\propto2\Lambda-\alpha\,g(x)$ phantom divide is crossed when  $g'(x)=0$.
This crossing is independent from $\alpha$ and $\mu$.
In particular,  our deformation gives rise to the following crossing points: $x_\pm=3\pm2\sqrt2$.

\item 
A regular background and an invertible Friedmann constraint restrict the deformation parameter $\alpha$ ti the interval $-1/3<\alpha<0.962$

\item We confront the background expansion with cosmic chronometers, DESI DR2 baryon acoustic oscillations (BAO),
Planck distance priors and three supernova (SN) compilations, and consider growth-rate data as an exploratory extension.
For the reference scale $\eta=\sqrt6$, chosen before looking at the data, the best fit has $\alpha>0$ for all three SN
compilations, with an improvement over \LCDM\ of $1.4$--$1.7\sigma$, while the Bayes factor mildly favours \LCDM. If the
data are allowed to choose $\eta$, the improvement grows to $\dchi\simeq4.8$--$6.9$ at $\eta=1.6$; Monte Carlo
simulations that repeat the search over $\eta$ give $2.1$--$2.5\sigma$. We also compare these results with other models: $w$CDM, with the
$w_0w_a$ parametrization and with two standard $f(Q)$ models on exactly the same data: the model gives the largest
improvement among the one-parameter extensions.
\end{itemize}
Throughout the paper we use natural units ($c=\hbar=1$), the mostly-plus signature, and denote by $\kappa\equiv8\pi G$
the gravitational coupling.

\section{Symmetric teleparallel setup and model definition}
\label{sec:setup}

\subsection{$f(Q)$ gravity}

In STEGR, curvature and torsion are zero, and the affine connection is not compatible with the metric. The quantity that encodes the latter property is the non-metricity tensor:

\begin{equation}
Q_{\alpha\mu\nu}\equiv\nabla_\alpha g_{\mu\nu}\,,
\label{eq:Qtensor}
\end{equation}
Its two independent traces are:
\begin{equation}
Q_\alpha\equiv Q_\alpha{}^\mu{}_\mu\,,\qquad \tilde Q_\alpha\equiv Q^\mu{}_{\alpha\mu}\,.
\label{eq:traces}
\end{equation}
Therefore the disformation tensor and the affine connection are
\begin{align}
L^\alpha{}_{\mu\nu}&=\frac12 g^{\alpha\beta}\left(-Q_{\mu\nu\beta}-Q_{\nu\mu\beta}+Q_{\beta\mu\nu}\right),\\
\Gamma^\alpha{}_{\mu\nu}&=\left\{{}^{\,\alpha}_{\mu\nu}\right\}+L^\alpha{}_{\mu\nu}\,,
\end{align}
with $\{{}^{\,\alpha}_{\mu\nu}\}$ the Levi-Civita symbols. The superpotential
\begin{align}
P^\alpha{}_{\mu\nu}={}&-\frac14 Q^\alpha{}_{\mu\nu}+\frac12 Q_{(\mu}{}^\alpha{}_{\nu)}\nonumber\\
&+\frac14\left(Q^\alpha-\tilde Q^\alpha\right)g_{\mu\nu}-\frac14\,\delta^\alpha_{(\mu}Q_{\nu)}
\end{align}
gives the STEGR scalar the compact form~\citep{BeltranJimenez2018CGR,BeltranJimenez2022Lost,Jensko2025,
BeltranJimenez2021Accidental}
\begin{equation}
Q=-Q_{\alpha\mu\nu}P^{\alpha\mu\nu}\,.
\label{eq:Qscalar}
\end{equation}

\subsection{Action and field equations}

We consider the nonlinear extension of STEGR
\begin{equation}
S\left[g_{\mu\nu},\Gamma^\alpha{}_{\mu\nu}\right]=-\frac{1}{2\kappa}\int d^4x\sqrt{-g}\,f(Q)+S_m[g,\Psi]\,.
\label{eq:action}
\end{equation}
Varying with respect to the metric gives, in the conventions of \citet{BeltranJimenez2020Cosmo},
\begin{align}
&\frac{2}{\sqrt{-g}}\nabla_\alpha\!\left(\sqrt{-g}\,f_Q P^\alpha{}_{\mu\nu}\right)+\frac12 g_{\mu\nu}f\nonumber\\
&\qquad+f_Q\left(P_{\mu\alpha\beta}Q_\nu{}^{\alpha\beta}-2Q_{\alpha\beta\mu}P^{\alpha\beta}{}_\nu\right)=\kappa T_{\mu\nu}\,,
\label{eq:fieldeq}
\end{align}
with $f_Q\equiv df/dQ$. 
If we consider an homogeneous and isotropic universe in the coincident gauge, the equation obtained by varying the connection is automatically satisfied,

\begin{equation}
\Gamma^\alpha{}_{\mu\nu}=0\,,\qquad Q_{\alpha\mu\nu}=\partial_\alpha g_{\mu\nu}\,.
\end{equation}
For a spatially flat FLRW metric, $g_{\mu\nu}={\rm diag}\left(-1,a^2,a^2,a^2\right)$, this gives
\begin{equation}
Q=6H^2\,,
\label{eq:Q6H2}
\end{equation}
and Eqs.~\eqref{eq:fieldeq} reduce to the two Friedmann equations~\citep{BeltranJimenez2020Cosmo}
\begin{align}
6f_QH^2-\frac12 f&=\kappa\left(\rho_m+\rho_r\right),\label{eq:fried1}\\
\left(12H^2f_{QQ}+f_Q\right)\dot H&=-\frac{\kappa}{2}\left(\rho_m+\rho_r+p_m+p_r\right),\label{eq:fried2}
\end{align}
together with the usual conservation equations $\dot\rho_i+3H(\rho_i+p_i)=0$ for matter and radiation. Once the matter
content is specified, Eq.~\eqref{eq:fried1} is algebraic in $H^2$, which makes the background analysis of $f(Q)$
cosmology particularly transparent~\citep{Heisenberg2024,BeltranJimenez2020Cosmo}.

The coincident gauge with Cartesian coordinates is one of the three connection branches compatible with a flat FLRW
metric~\citep{Dimakis2022,Jensko2025}. Equations~\eqref{eq:fried1} and \eqref{eq:fried2} are those of this branch. \citet{Gomes2024} have shown that, for nonlinear $f(Q)$, perturbations around FLRW are infinitely strongly coupled in
this branch and in one of the other two, while the remaining branch propagates a ghost. In the following we therefore base our results on the
background; the quasi-static growth of matter perturbations is considered in Sec.~\ref{sec:growth} as an exploratory
extension, and the implications are discussed in Sec.~\ref{sec:discussion}.

\subsection{The Hu--Sawicki-inspired model}

The deformation in Eq.~\eqref{eq:model} follows the rationale of the Hu--Sawicki model of $f(R)$
gravity~\citep{HuSawicki2007}, with the STEGR scalar $Q$ as geometric variable, although its functional form is
different. Unlike the Hu--Sawicki model, whose form was chosen so that the chameleon mechanism hides the modification in
the Solar System, no screening mechanism is known for $f(Q)$ gravity (Sec.~\ref{sec:growth}). The deformation is
designed to satisfy three requirements:
\begin{enumerate}
\item the deformation term in the Lagrangian vanishes at $Q=\mu^2$, i.e.\ at $x=1$, which provides a reference scale;
\item at high curvature the deformation term is suppressed as $1/Q$, so that the model approaches STEGR$+\Lambda$ in the
early Universe, preserving the successes of nucleosynthesis and of the CMB;
\item at low curvature it approaches a constant, which acts as a shift of the cosmological constant.
\end{enumerate}
It is convenient to introduce the dimensionless ratios
\begin{equation}
\eta\equiv\frac{\mu}{H_0}\,,\qquad x_0\equiv\frac{Q_0}{\mu^2}=\frac{6}{\eta^2}\,,
\label{eq:eta}
\end{equation}
which fix the redshift at which the deformation becomes relevant. Our reference choice is
\begin{equation}
\eta=\sqrt6\quad\Longleftrightarrow\quad\mu^2=Q_0=6H_0^2\,,\quad x_0=1\,,
\label{eq:etasqrt6}
\end{equation}
which is fixed before any comparison with the data. For this value the deformation term in the Lagrangian vanishes at
the present epoch. This does not make the model coincide with \LCDM\ today: $f_Q(x_0)=1+\alpha/4\neq1$, so the effective
gravitational coupling and the dark-energy equation of state at $z=0$ differ from their \LCDM\ values
(Secs.~\ref{sec:friedmann} and \ref{sec:growth}). The choice is a convention: a constant term in $f(Q)$ can be
absorbed into $\Lambda$, so the vanishing of the deformation today does not single out $\mu^2=Q_0$ physically. The
dependence of the results on $\eta$, and the possibility of determining it from the data, are discussed in
Sec.~\ref{sec:eta}.

\section{Modified Friedmann equations and effective dark energy}
\label{sec:friedmann}

\subsection{Dimensionless Friedmann constraint}

Substituting Eq.~\eqref{eq:model} into Eq.~\eqref{eq:fried1} and separating the deformation term, the Friedmann
constraint takes the compact form
\begin{equation}
x-R(x)=x_0\left[\Om(1+z)^3+\Or(1+z)^4\right],
\label{eq:constraint}
\end{equation}
where
\begin{equation}
R(x)\equiv\frac{2\kappa\,\rho_{\rm DE}(x)}{\mu^2}=2\Lambda-\alpha\,\frac{1+6x-3x^2}{(1+x)^3}
\label{eq:R}
\end{equation}
is the effective dark-energy density in units of $\mu^2/(2\kappa)$, and $\Lambda$ is expressed in units of $\mu^2$.
Imposing $x=x_0$ at $z=0$ fixes
\begin{equation}
2\Lambda=\left(1-\Om-\Or\right)x_0+\alpha\,\frac{1+6x_0-3x_0^2}{(1+x_0)^3}\,.
\label{eq:Lambda}
\end{equation}
The radiation term is negligible at late times but is needed for the CMB distance priors and for the sound horizon
(Sec.~\ref{sec:data}). The other background quantities follow by direct substitution:
\begin{align}
f_Q(x)&=1+\alpha\,\frac{3-x}{(1+x)^3}\,,\label{eq:fQ}\\
f_{QQ}(x)&=\frac{2\alpha\,(x-5)}{\mu^2(1+x)^4}\,,\label{eq:fQQ}\\
D(x)&\equiv f_Q+2Qf_{QQ}=1-R'(x)\nonumber\\&=1+3\alpha\,\frac{x^2-6x+1}{(1+x)^4}\,.\label{eq:D}
\end{align}
Differentiating Eq.~\eqref{eq:constraint} with respect to $N\equiv\ln a$ gives
\begin{equation}
\frac{dx}{dN}=-\frac{x_0\left[3\Om(1+z)^3+4\Or(1+z)^4\right]}{D(x)}\,,
\label{eq:dxdN}
\end{equation}
which is equivalent to Eq.~\eqref{eq:fried2}. As long as $D>0$, Eq.~\eqref{eq:constraint} has a unique solution
$x(z)$, which increases monotonically with $z$; we obtain it numerically with a Newton iteration on the branch that
passes through $x_0$ today. Interpreting $R$ as the density of an effective fluid, its equation of state is
\begin{equation}
w_{\rm DE}=-1-\frac13\frac{d\ln R}{dN}=-1-\frac{R'(x)}{3R(x)}\,\frac{dx}{dN}\,.
\label{eq:wde}
\end{equation}

\subsection{Range of background regularity}
\label{sec:prior}

The background is regular only if $f_Q>0$, so that the effective gravitational coupling $G_{\rm eff}=G/f_Q$
(Sec.~\ref{sec:growth}) is positive, and $D>0$, so that Eq.~\eqref{eq:constraint} can be inverted and $\dot H$ stays
finite. Requiring both conditions for every $x>0$, i.e.\ for every value of $Q$ and not only along a particular
cosmological trajectory, gives
\begin{equation}
-\frac13<\alpha<\frac{1}{3\,|q_{\rm min}|}\simeq0.962\,,\qquad q(x)\equiv\frac{x^2-6x+1}{(1+x)^4}\,,
\label{eq:alpharange}
\end{equation}
where $q(0)=1$ and $q$ reaches its minimum $q_{\rm min}\simeq-0.347$ at $x=5-2\sqrt5$. The condition on $f_Q$ gives the
same lower bound and the weaker upper bound $\alpha<108$. These conditions make the
background regular for every $x>0$, but they do not guarantee a future de Sitter point. Since $F(x)\equiv x-R(x)$ has
$F'(x)=D(x)>0$, a de Sitter point with $x>0$ exists if and only if $F(0)<0$, i.e.\ $R(0)>0$:
\begin{equation}
x_0\left(1-\Om-\Or\right)+\alpha\left[\frac{1+6x_0-3x_0^2}{(1+x_0)^3}-1\right]>0\,.
\label{eq:desitter}
\end{equation}
Otherwise $H$ vanishes at a finite matter density and the expansion turns around (for instance at $a\simeq6.6$ for
$\eta=2.1$, $\alpha=0.95$, $\Om=0.495$). Condition~\eqref{eq:desitter} removes a corner of the prior volume ($\alpha>0.95$,
$\Om>0.49$, $2.07<\eta<2.27$; a fraction $2\times10^{-5}$), far from the posterior. The prior on $\alpha$ used throughout
this work (Table~\ref{tab:priors}) is Eq.~\eqref{eq:alpharange} intersected with condition~\eqref{eq:desitter}: every
sampled model has a regular past, from the radiation era, and a regular future up to the de Sitter point. Requiring
regularity only along the cosmological trajectory, rather than for every $Q$, would allow larger values of $\alpha$ for
$\eta\lesssim1.5$ (Sec.~\ref{sec:eta}).

\subsection{Phantom crossing}

\begin{figure*}
\centering
\includegraphics[width=0.92\textwidth]{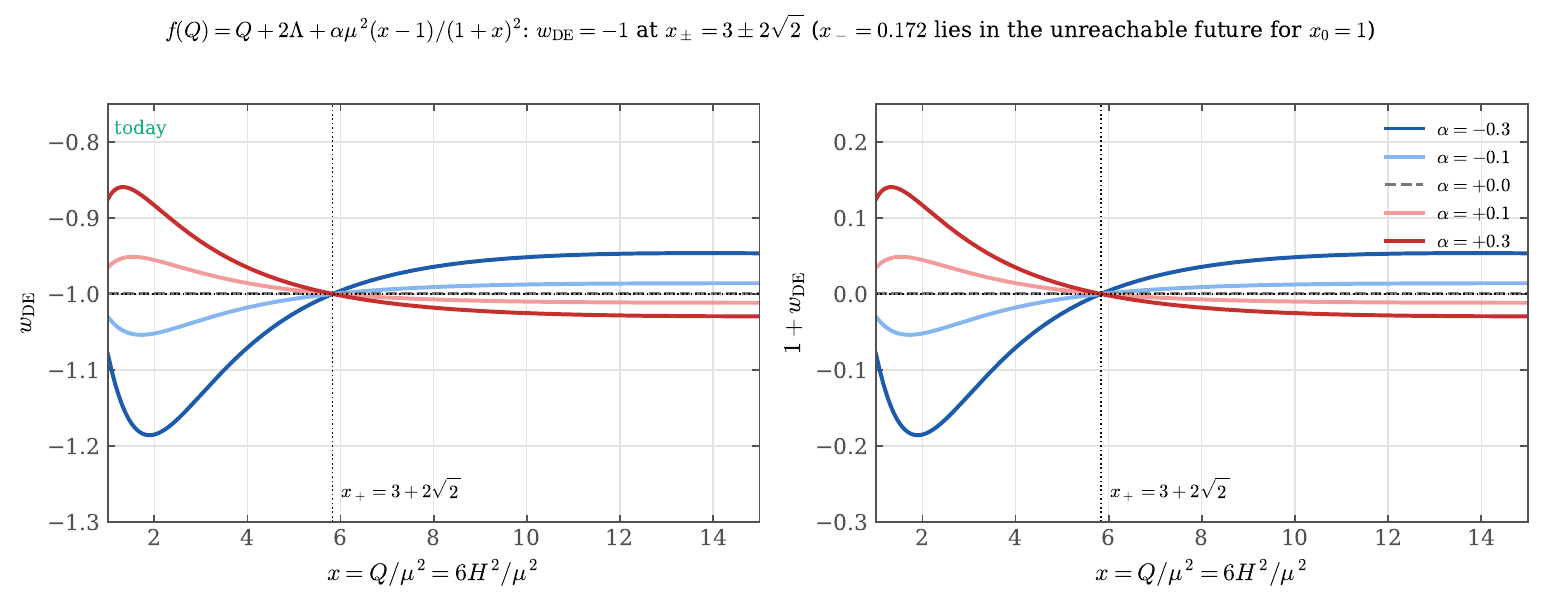}
\caption{Structural origin of the phantom crossing. Left: effective dark-energy equation of state $w_{\rm DE}(x)$ for
several values of $\alpha$, along the past light cone $x\geq x_0=1$ ($\eta=\sqrt6$) of a matter-dominated background with
$\Om=0.3$; the left edge of the plot is the present epoch. Right: $1+w_{\rm DE}(x)$ for the same parameters. For every
$\alpha\neq0$, $w_{\rm DE}$ crosses $-1$ at $x_+=3+2\sqrt2$ (dotted line). The lower root $x_-=0.172$ lies below the
future de Sitter point and is never reached for $x_0=1$.}
\label{fig:wdex}
\end{figure*}

Crossing the phantom divide is a well-known difficulty for single-field models of dark energy~\citep{Vikman2005} and
has motivated a broad literature on dynamical dark energy, from quintessence~\citep{Caldwell1998} to
phantom~\citep{Caldwell2002} and quintom~\citep{Feng2005,Guo2005,Cai2010} scenarios (see \citealt{Copeland2006} for a
review). As for any deformation linear in $\alpha$, the phantom divide is crossed exactly where
$R'(x)=0$~\citep[see also][]{AmendolaTsujikawa2008,WuYu2011}. From Eq.~\eqref{eq:R},
\begin{equation}
R'(x)=-3\alpha\,\frac{x^2-6x+1}{(1+x)^4}\,,
\end{equation}
so that the crossing roots are
\begin{equation}
x_\pm=3\pm2\sqrt2\simeq\{0.172,\,5.828\}\,.
\label{eq:roots}
\end{equation}
Both roots are independent of $\alpha$ (provided $\alpha\neq0$) and of $\mu$. Figure~\ref{fig:wdex} shows that, for all
values of $\alpha$, $w_{\rm DE}(x)$ crosses $-1$ at the same value $x_+$, while the amplitude and the sign of the
deviation on either side are controlled by $\alpha$. For $\alpha>0$ the equation of state is phantom before the crossing
and quintessence-like after it, the direction found in the DESI $w_0w_a$ fits~\citep{DESI2025DR2}.

What is structural is the location of the crossing in $x$, not its redshift. Along the past trajectory $x\geq x_0$,
so the upper root is crossed in the past for $x_0<x_+$, i.e.\ $\eta>1.01$, while the lower root is crossed in the past
only for $x_0<x_-$, i.e.\ $\eta>5.9$, outside the range considered here. For $1.01<\eta<5.9$ the cosmological
trajectory therefore crosses the phantom divide once, at
\begin{equation}
x_0\,E^2(z_{\rm cross})=x_+\,,
\end{equation}
where $E=H/H_0$. For the best-fit parameters of the main analysis at $\eta=\sqrt6$ (Sec.~\ref{sec:results}) this gives
$z_{\rm cross}=1.55$ with Pantheon+, $1.54$ with DES-Y5 and $1.53$ with Union3; at $\eta=1.6$ it gives $z_{\rm cross}=0.77$--$0.78$. The
crossing redshift is therefore controlled by $\eta$ and, more weakly, by $\Om$.

\section{Background-level phenomenology}
\label{sec:pheno}

\begin{figure*}
\centering
\includegraphics[width=0.92\textwidth]{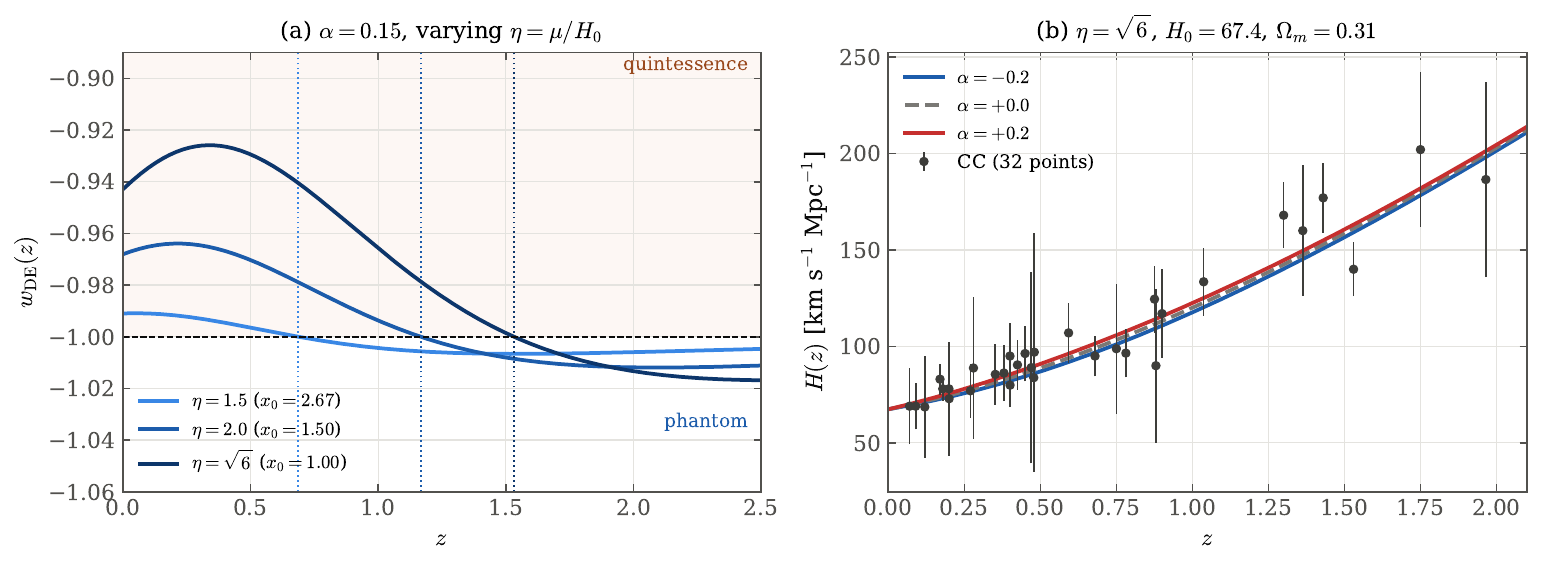}
\caption{Background phenomenology. (a) Effective equation of state $w_{\rm DE}(z)$ for $\alpha=0.15$ and three values of
$\eta=\mu/H_0$; the dotted lines mark the crossing redshifts, which move with $\eta$. (b) $H(z)$ for $\alpha=-0.2,0,+0.2$
at $\eta=\sqrt6$, $H_0=67.4$ and $\Om=0.31$, compared with the 32 cosmic-chronometer measurements; within the CC
uncertainties the three curves are indistinguishable. The comparison of the posterior $H(z)$ with the CC data is shown in
Fig.~\ref{fig:hz}.}
\label{fig:background}
\end{figure*}

Three consequences follow from the structure described above (Fig.~\ref{fig:background}). First, varying $\eta$ at fixed
$\alpha$ mainly changes the redshift of the phantom crossing, while leaving the qualitative evolution unchanged. For
$\alpha>0$ the effective equation of state is quintessence-like near the present epoch, crosses the phantom divide at
$z_{\rm cross}$, and then relaxes back towards $w_{\rm DE}=-1$ at high redshift. Second, $\alpha$ has a small effect on
the expansion rate over the redshift range probed by cosmic chronometers (CC): for the best fits of
Sec.~\ref{sec:results} $H(z)$ differs from \LCDM\ by less than $1.7\%$ at $z<0.25$ and by less than $0.6\%$ elsewhere,
well below the CC uncertainties. Third, the BAO distance ratios also stay close to \LCDM; the constraint on $\alpha$
arises from the combination of SN, BAO and CMB data rather than from any single probe (Table~\ref{tab:perprobe}).

The model thus keeps the background expansion close to \LCDM\ while modifying the dark-energy equation of state in a
controlled way, which makes it relevant for the discussion of evolving dark energy prompted by the DESI results.

\section{Data sets and statistical methodology}
\label{sec:data}

\subsection{Data}

The analysis combines the data sets summarised in Table~\ref{tab:data}. We describe below the main choices entering
the likelihood.

\paragraph{Cosmic chronometers} We use 32 measurements of $H(z)$ in $z\in[0.07,1.97]$, obtained from the differential
age of passively evolving galaxies~\citep{Stern2010,Moresco2012,Moresco2016,Moresco2022}, in the compilation of
\citet{Favale2023}. We use diagonal errors that include statistical and systematic contributions. The systematic
errors from stellar population synthesis are correlated~\citep{Moresco2020}; since the CC data carry little weight in the
comparison between models ($|\dchi_{\rm CC}|\lesssim0.2$, Table~\ref{tab:perprobe}), neglecting these correlations does
not affect our conclusions.

\paragraph{DESI DR2 BAO} We use the 13 BAO measurements of DESI DR2~\citep{DESI2025DR2} with their full covariance:
$D_V/r_d$ for the BGS sample at $z=0.295$, and $D_M/r_d$ and $D_H/r_d$ for LRG1, LRG2, LRG3+ELG1, ELG2, QSO and
Ly$\alpha$, in $z\in[0.295,2.33]$. The sound horizon at the drag epoch, $r_d$, is computed at each step from $\omega_m$
and $\omega_b$ with the calibrated formula of \citet{Aubourg2015}, including a neutrino mass
$\sum m_\nu=0.06$~eV, which is accurate to $0.1\%$ (we find agreement with CAMB~\citep{Lewis2000CAMB} to $0.03\%$
over the posterior range); the Eisenstein--Hu fitting formula~\citep{EisensteinHu1998} has
errors of $1$--$2\%$, larger than the precision of DR2.

\paragraph{Compressed Planck 2018 likelihood} For the CMB we use the distance priors derived by \citet{Chen2019} from Planck 2018 TT,TE,EE+lowE~\citep{Planck2018VI}: the shift parameter
$R=\sqrt{\Om H_0^2}\,D_M(z_*)=1.7502\pm0.0046$, the acoustic scale $\ell_A=\pi D_M(z_*)/r_s(z_*)=301.471\pm0.0895$ and
$\omega_b=0.02236\pm0.00015$, with their correlation matrix. The comoving sound horizon $r_s(z_*)$ is integrated
numerically, with $\Omega_r=\omega_r/h^2$, $\omega_r=\omega_m/(1+z_{\rm eq})$ and
$z_{\rm eq}=2.5\times10^4\,\omega_m(T_{\rm CMB}/2.7\,{\rm K})^{-4}$, and $z_*$ is computed at each step from $\omega_b$ and
$\omega_m$ with the fitting formula of \citet{HuSugiyama1996}, shifted by $-1.999$ to match CAMB~\citep{Lewis2000CAMB}
(rms error $0.04$ for $0.0218<\omega_b<0.0230$ and $0.135<\omega_m<0.150$). This is needed because \citet{Chen2019}
built the priors with the $z_*$ of each sample of the Planck chains: a fixed $z_*=1089.80$ would overestimate $\ell_A$
by $0.05$--$0.08$ ($0.6$--$0.9\sigma$) at our best fits, although $\dchi$ changes by less than $0.05$
(\ref{app:robust} and \ref{app:cmb}). The compressed likelihood is a reasonable first approximation because the
deformation decays as $1/x$ at early times, but it does not capture the late-time integrated Sachs--Wolfe effect or CMB
lensing, which the deformation also modifies; a full analysis of the CMB anisotropies would require a Boltzmann
implementation of the model, which is beyond the scope of this work.

\paragraph{Pantheon+} Our baseline SN sample is the Pantheon+ compilation~\citep{Scolnic2022,Brout2022}, restricted to
$z_{\rm HD}>0.01$ to suppress peculiar-velocity systematics. This leaves $N_{\rm SN}=1590$ light curves in
$z\in[0.010,2.26]$, for which we use the full statistical-plus-systematic covariance $C_{\rm SN}$. Luminosity distances
are computed as $D_L=(1+z_{\rm hel})\,D_M(z_{\rm HD})$. The absolute magnitude $M_B$ is degenerate with the distance
scale and is marginalised analytically~\citep{Goliath2001,Conley2011},
\begin{equation}
\chi^2_{\rm SN}=\mathbf r^{\rm T}C_{\rm SN}^{-1}\mathbf r-\frac{\left(\mathbf r^{\rm T}C_{\rm SN}^{-1}\mathbf 1\right)^2}
{\mathbf 1^{\rm T}C_{\rm SN}^{-1}\mathbf 1}\,,
\label{eq:chi2sn}
\end{equation}
with $r_i=m_{B,i}^{\rm obs}-\mu_{\rm th}(z_i;\bm\theta)$ and $\mathbf 1$ a vector of ones.

\paragraph{DES-Y5} We use the five-year SN sample of the Dark Energy Survey~\citep{DES2024} in the version recalibrated
with the Dovekie photometric calibration~\citep{Popovic2026}: 1820 SNe in $z\in[0.025,1.14]$, with the full
statistical-plus-systematic covariance. With this recalibration the preference of DESI+CMB+DES SNe for evolving dark
energy decreases from $4.2\sigma$ to $3.2\sigma$~\citep{Popovic2026}. $M_B$ is marginalised as in
Eq.~\eqref{eq:chi2sn}.

\paragraph{Union3} We use the 22 binned distance moduli of Union3~\citep{Rubin2025} in $z\in[0.05,2.26]$ with their
full covariance; the offset is marginalised as in Eq.~\eqref{eq:chi2sn}.

\paragraph{Growth rate} In the exploratory extension of Sec.~\ref{sec:growth} we use the Gold-2017 compilation of 18 measurements of $f\sigma_8(z)$ in
$z\in[0.02,1.4]$~\citep{Nesseris2017}, with the published covariance of the three WiggleZ points. The dependence of each
measurement on the fiducial cosmology assumed in the original analysis is corrected at the level of the theory
prediction~\citep{Nesseris2017},
\begin{equation}
f\sigma_8^{\rm th}(z)\;\to\;f\sigma_8^{\rm th}(z)\,\frac{H(z)D_A(z)}{\left[H(z)D_A(z)\right]_{\rm fid}}\,,
\end{equation}
where the fiducial quantity is computed in flat \LCDM\ with the value of $\Omega_m$ used for each measurement.

\begin{table*}
\centering
\caption{Data sets entering the joint analysis. ``Size'' gives the number of data points; ``Cov.'' indicates the
treatment of correlations; ``Nuisance'' lists the parameters marginalised analytically. The main analysis combines CC,
BAO, CMB and one SN compilation, with $N=1638$, $1868$ and $70$ data points for Pantheon+, DES-Y5 and Union3,
respectively; the growth data enter only the extension of Sec.~\ref{sec:growth}.}
\label{tab:data}
\footnotesize
\begin{adjustbox}{max width=\linewidth}
\begin{tabular}{llcccl>{\raggedright\arraybackslash}p{3.6cm}}
\toprule
Data set & Observable & Size & $z$ range & Cov. & Nuisance & Ref. \\
\midrule
Cosmic chronometers & $H(z)$ & 32 & $[0.07,1.97]$ & diagonal & -- & \citealp{Favale2023} \\
DESI DR2 BAO & $D_{V,M,H}/r_d$ & 13 & $[0.295,2.33]$ & full & -- & \citealp{DESI2025DR2} \\
Planck 2018 priors & $R$, $\ell_A$, $\omega_b$ & 3 & $z_*(\omega_b,\omega_m)$ & full & -- & \citealp{Chen2019,Planck2018VI} \\
Pantheon+ & $m_B$ & 1590 & $[0.010,2.26]$ & stat+sys & $M_B$ & \citealp{Scolnic2022,Brout2022} \\
DES-Y5 (Dovekie) & $\mu$ & 1820 & $[0.025,1.14]$ & stat+sys & $M_B$ & \citealp{DES2024,Popovic2026} \\
Union3 (binned) & $\mu$ & 22 & $[0.05,2.26]$ & full & offset & \citealp{Rubin2025} \\
Growth (Gold-2017)$^\dagger$ & $f\sigma_8(z)$ & 18 & $[0.02,1.4]$ & WiggleZ block & -- & \citealp{Nesseris2017} \\
\bottomrule
\end{tabular}
\end{adjustbox}
\begin{flushleft}\footnotesize $^\dagger$Only in the exploratory extension of Sec.~\ref{sec:growth}.\end{flushleft}
\end{table*}

\subsection{Likelihood and sampling}

For a parameter vector $\bm\theta$ the likelihood is Gaussian,
\begin{equation}
\chi^2(\bm\theta)=\sum_{\rm probes}\left[\mathcal O^{\rm obs}-\mathcal O^{\rm th}(\bm\theta)\right]^{\rm T}C^{-1}
\left[\mathcal O^{\rm obs}-\mathcal O^{\rm th}(\bm\theta)\right],
\end{equation}
so that the posterior is $\mathcal P(\bm\theta|{\rm data})\propto\pi(\bm\theta)\,e^{-\chi^2/2}$, with $\pi(\bm\theta)$ the
prior. At fixed $\eta$ we sample the four-dimensional vector
\begin{equation}
\bm\theta=(\alpha,\,\Om,\,H_0,\,\omega_b)\,,
\end{equation}
add $\eta$ as a fifth parameter in the analysis of Sec.~\ref{sec:eta}, and $\sigma_{8,0}$ in the extension of
Sec.~\ref{sec:growth}. The priors are uniform and listed in Table~\ref{tab:priors}; the prior on $\alpha$ is the
regularity range of Eq.~\eqref{eq:alpharange}, with condition~\eqref{eq:desitter}.

Our baseline sampler is an adaptive Metropolis--Hastings (MH) algorithm. The proposal is initialised as a multivariate
Gaussian with diagonal widths $(\sigma_\alpha,\sigma_{\Omega_m},\sigma_{H_0},\sigma_{\omega_b})=
(0.05,0.006,0.6,1.5\times10^{-4})$, with $\sigma_{\sigma_8}=0.02$ when $\sigma_{8,0}$ is sampled. During burn-in the overall scale is adjusted to keep the acceptance rate between
$0.22$ and $0.40$; the covariance is estimated from the second half of the burn-in and kept fixed during production.
For each data combination we run four chains from overdispersed starting points (Table~\ref{tab:runs}). Convergence is
assessed with the rank-normalised split $\hat R$~\citep{Gelman1992,Vehtari2021} and with the bulk and tail
effective sample sizes of \citet{Vehtari2021}: all fixed-$\eta$ runs satisfy $\hat R<1.006$ and have effective sample sizes larger than 900
for every parameter. As a cross-check we use the affine-invariant ensemble sampler {\tt emcee}~\citep{Goodman2010,
ForemanMackey2013} with the stretch move, for which we require chains longer than $50$ integrated autocorrelation times (the run with $\eta$ free reaches
$61$ for $\eta$, see Sec.~\ref{sec:eta}).
In the analysis with $\eta$ free the MH sampler does not mix along the ridge $\alpha\simeq0$, where
$\eta$ is unconstrained, and we use {\tt emcee} (Sec.~\ref{sec:eta}).

\begin{table}
\centering
\caption{MCMC runs. The fixed-$\eta$ runs are repeated for each SN compilation. Acceptance rates are given per chain
for MH and as the mean over walkers for {\tt emcee}; for {\tt emcee} the number of chains is the number of walkers.}
\label{tab:runs}
\small
\begin{adjustbox}{max width=\linewidth}
\begin{tabular}{lccccc}
\toprule
Run & Sampler & $N_{\rm dim}$ & Chains & $N_{\rm burn}$ & $N_{\rm prod}$ \\
\midrule
$\eta=\sqrt6$ & MH & 4 & 4 & 4000 & 20000 \\
$\eta=1.6$ & MH & 4 & 4 & 4000 & 50000 \\
$\eta$ free (P+) & {\tt emcee} & 5 & 32 & 1000 & 40000 \\
Check, $\eta=\sqrt6$ & {\tt emcee} & 4 & 32 & 1000 & 4000 \\
\bottomrule
\end{tabular}
\end{adjustbox}
\begin{flushleft}\footnotesize
Acceptance: $0.26$--$0.31$ ($\eta=\sqrt6$), $0.26$--$0.32$ ($\eta=1.6$), $0.34$ ($\eta$ free), $0.59$ (cross-check).
At $\eta=1.6$ the posterior of $\alpha$ reaches the regularity bound and longer chains are needed.
\end{flushleft}
\end{table}

\begin{table}
\centering
\caption{Uniform priors. $\eta$ is varied only in the analysis of Sec.~\ref{sec:eta}.}
\label{tab:priors}
\begin{adjustbox}{max width=\linewidth}
\begin{tabular}{lc}
\toprule
Parameter & Prior \\
\midrule
$\alpha$ & $\mathcal U(-1/3,\,0.962)$ [Eqs.~\eqref{eq:alpharange} and \eqref{eq:desitter}] \\
$\Om$ & $\mathcal U(0.10,\,0.50)$ \\
$H_0$ [km\,s$^{-1}$\,Mpc$^{-1}$] & $\mathcal U(55,\,85)$ \\
$\omega_b$ & $\mathcal U(0.018,\,0.026)$ \\
$\sigma_{8,0}$ (Sec.~\ref{sec:growth} only) & $\mathcal U(0.5,\,1.1)$ \\
$\eta$ & $\mathcal U(1.0,\,3.5)$ \\
\bottomrule
\end{tabular}
\end{adjustbox}
\end{table}

\subsection{Model comparison}

We quantify the improvement over \LCDM\ with $\dchi\equiv\chi^2_{\Lambda,\rm min}-\chi^2_{f,\rm min}$, so that positive
values favour the deformation, and with the information criteria~\citep{Akaike1974,Schwarz1978,Liddle2007}
\begin{equation}
\Delta{\rm AIC}=\dchi-2\Delta k\,,\qquad \Delta{\rm BIC}=\dchi-\Delta k\ln N\,,
\label{eq:ic}
\end{equation}
where $\Delta k$ is the number of additional parameters and $N$ the total number of data points of the combination
(not the size of the SN sample alone). The $\ln N$ penalty therefore depends on how the SN data are represented
(22 Union3 nodes against about 1600--1800 light curves), and we compare $\Delta$BIC only within a given compilation.
At fixed $\eta$, $\Delta k=1$ and the significance of the improvement is $\sqrt{\dchi}$ in Gaussian units. When $\eta$ is
chosen from the same data it is counted as a fitted parameter in the information criteria ($\Delta k=2$). Its
significance, however, does not follow from a $\chi^2$ distribution with two degrees of freedom: since $\eta$ is not
defined when $\alpha=0$, Wilks' theorem does not apply to the maximum of $\dchi$ over
$\eta$~\citep{Davies1987,GrossVitells2010}. We calibrate it with Monte Carlo simulations (Sec.~\ref{sec:lee}). AIC
and BIC are information criteria and not Bayesian evidences. As a Bayesian measure we compute the Bayes factor of \LCDM\ against the model with
the Savage--Dickey density ratio~\citep{Trotta2007,Trotta2008},
\begin{equation}
B_{\Lambda f}=\frac{\mathcal P(\alpha=0\,|\,{\rm data})}{\pi(\alpha=0)}\,,
\label{eq:sddr}
\end{equation}
with the marginal posterior estimated by a kernel density and $\pi$ the uniform prior of Eq.~\eqref{eq:alpharange};
$B_{\Lambda f}>1$ favours \LCDM. The same ratio applies with $\eta$ free, since $\alpha=0$ is nested for every $\eta$.

\subsection{Significance with $\eta$ chosen from the data}
\label{sec:lee}

To calibrate the significance of the maximum of $\dchi$ over $(\alpha,\eta)$ we generate $2\times10^4$ mock data sets
from the \LCDM\ best fit of each combination, with Gaussian noise drawn from the full covariance of every probe, and
repeat on each mock the whole search: $\alpha$ on 120 nodes inside the prior and $\eta$ on the grid
$1.0,1.1,\dots,3.5$ of Sec.~\ref{sec:eta}, plus $\eta=\sqrt6$. At every node the nuisance parameters ($\Om$, $H_0$,
$\omega_b$ and the SN offset) are profiled to linear order around their best fit to the noise-free \LCDM\ data vector,
so that $\dchi$ is a projection in whitened data space; the dependence on $\alpha$ and $\eta$ is kept exact. Applied to
the real data, the procedure reproduces the profile of Sec.~\ref{sec:eta} to $0.01$ in $\dchi$. At fixed $\eta=\sqrt6$
the simulated distribution of $\dchi$ coincides with a $\chi^2$ distribution with one degree of freedom, as expected;
the distribution of the maximum over $\eta$ lies between those with one and two degrees of freedom (\ref{app:lee}).

\section{Main cosmological constraints ($\eta=\sqrt6$)}
\label{sec:results}

\begin{figure*}
\centering
\includegraphics[width=0.80\textwidth]{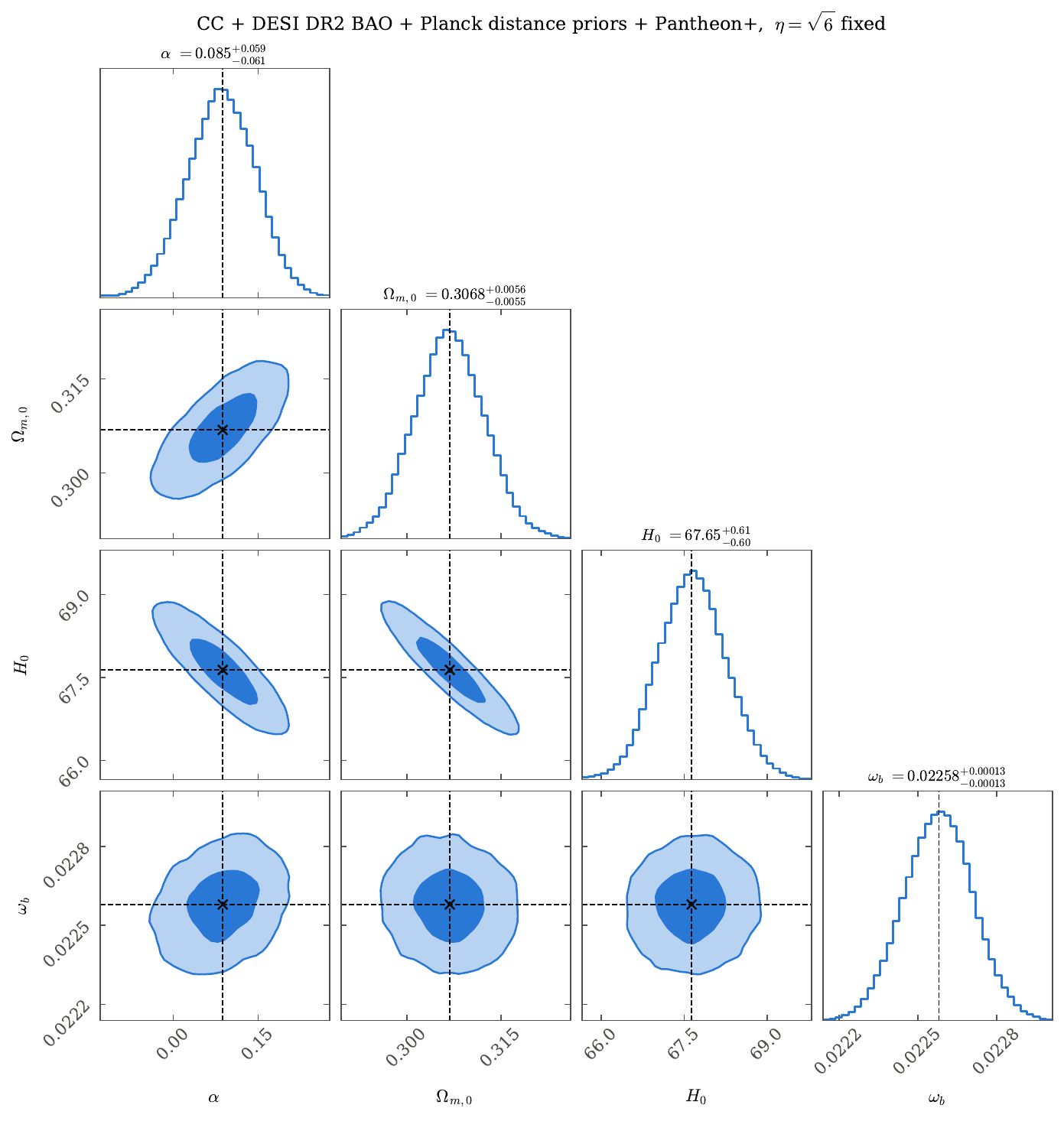}
\caption{Marginalised posterior of the baseline four-parameter analysis, CC + DESI DR2 BAO + Planck distance priors +
Pantheon+, at $\eta=\sqrt6$. Contours enclose $39.3\%$ and $86.5\%$ of the probability ($1\sigma$ and
$2\sigma$ in two dimensions); titles give the median and the $16$--$84\%$ interval; dashed lines mark the best fit.}
\label{fig:corner_p}
\end{figure*}

\begin{figure*}
\centering
\includegraphics[width=0.49\textwidth]{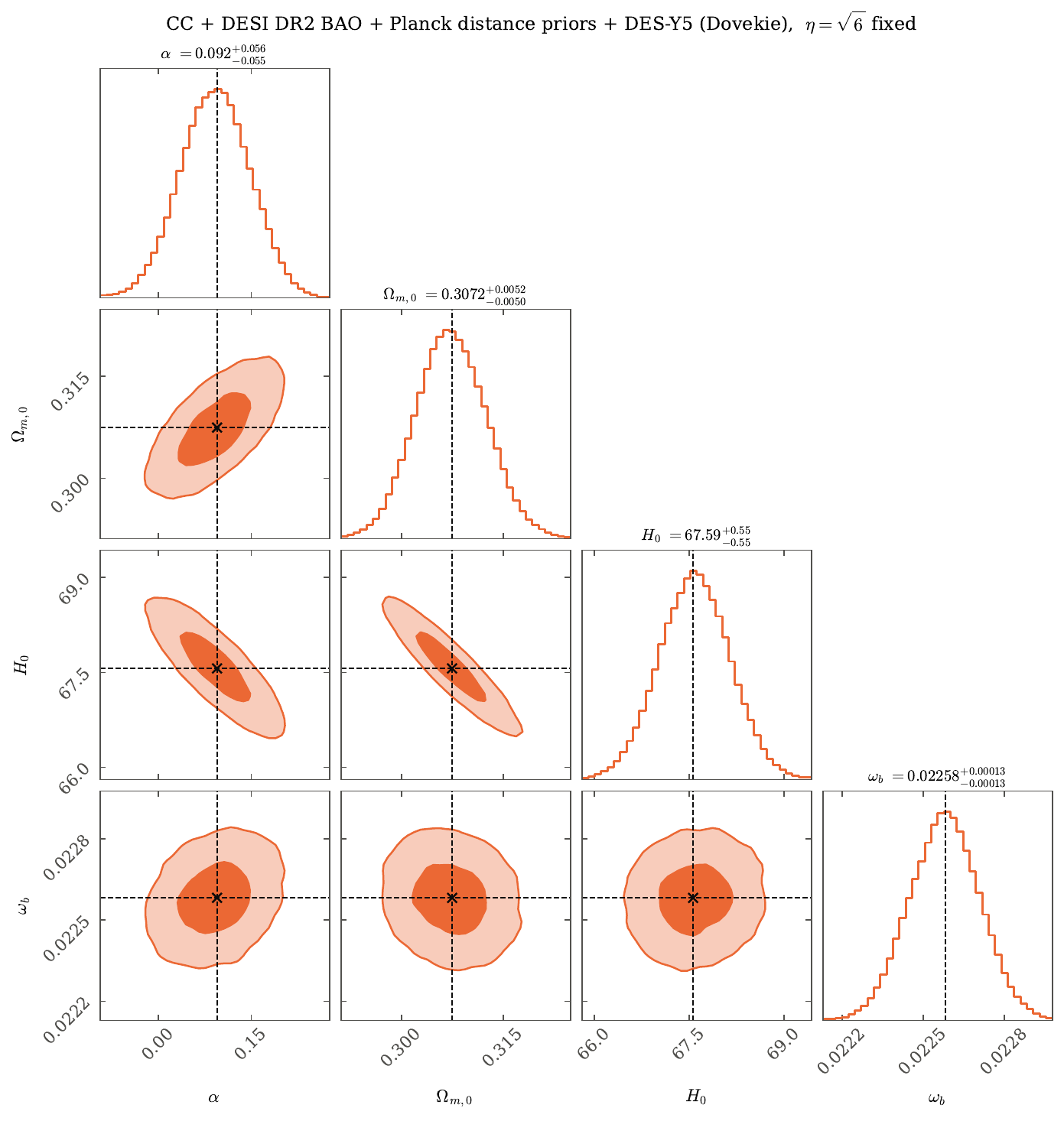}\hfill
\includegraphics[width=0.49\textwidth]{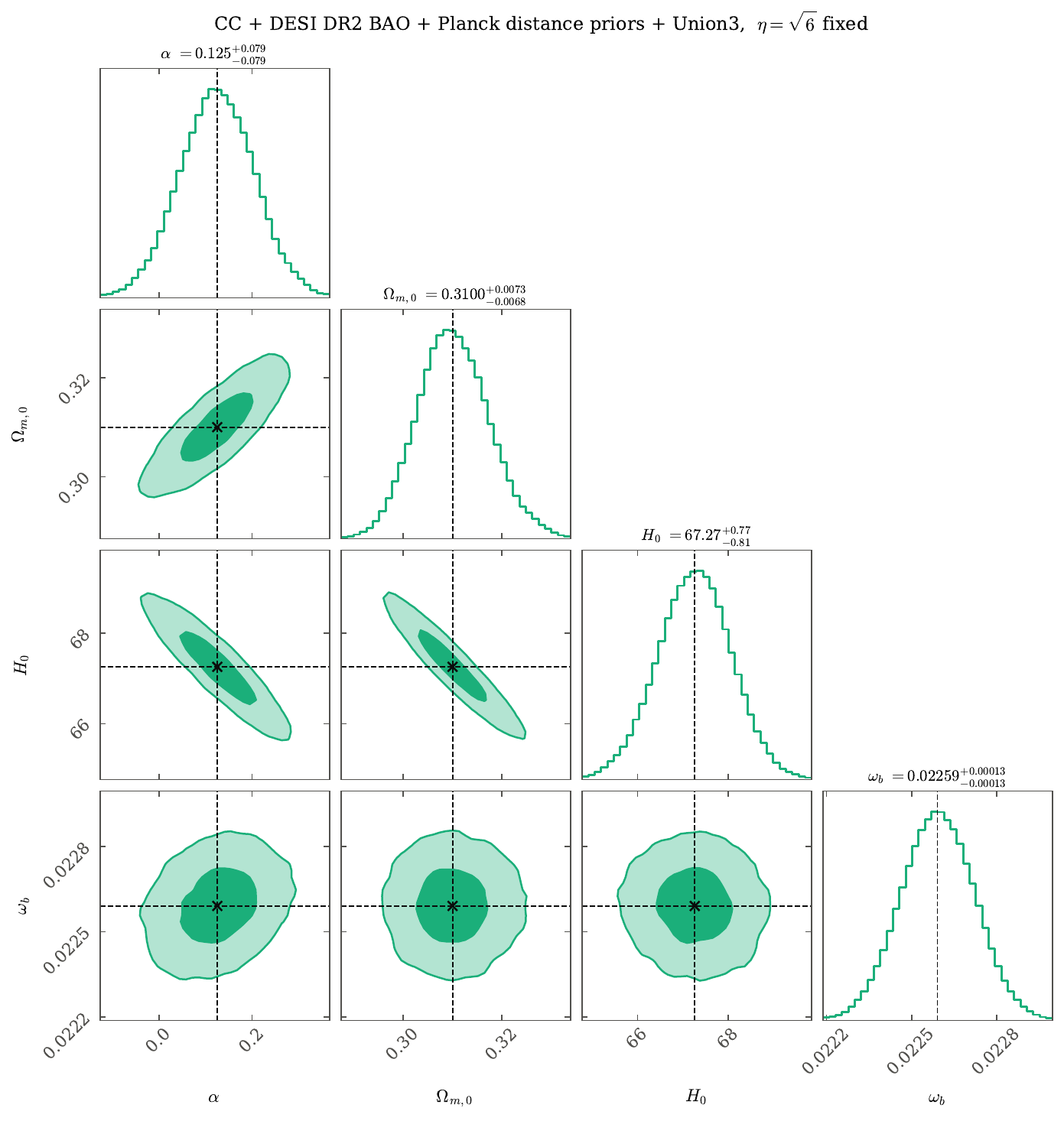}
\caption{As Fig.~\ref{fig:corner_p}, with DES-Y5 (Dovekie, left) and Union3 (right) in place of Pantheon+.}
\label{fig:corner_du}
\end{figure*}

Figure~\ref{fig:corner_p} shows the marginalised posterior of the baseline analysis, CC + DESI DR2 BAO + Planck distance
priors + Pantheon+, at $\eta=\sqrt6$. The deformation parameter is
\begin{equation}
\alpha=0.085\pm0.060\quad(\text{Pantheon+, mean}\pm1\sigma)\,,
\end{equation}
with best fit $\alpha=0.087$. The posterior is Gaussian and far from the prior boundaries. As expected, $\alpha$ is
positively correlated with $\Om$ and negatively correlated with $H_0$, while $\omega_b$, fixed by the CMB priors, is nearly insensitive to the deformation. The results for DES-Y5 and Union3 are
shown in Fig.~\ref{fig:corner_du} and summarised, together with the baseline, in Tables~\ref{tab:alpha}--\ref{tab:ic}.

\begin{table}
\centering
\caption{Deformation parameter at $\eta=\sqrt6$ for the three SN compilations. $\alpha$ is the marginalised mean and
standard deviation; $\sqrt{\dchi}$ is the significance of the improvement for one additional parameter (the
posterior-based estimate $|\langle\alpha\rangle|/\sigma$ gives $1.40$, $1.65$ and $1.57$); $B_{\Lambda f}$ is the
Savage--Dickey Bayes factor of Eq.~\eqref{eq:sddr} ($B_{\Lambda f}>1$ favours \LCDM).}
\label{tab:alpha}
\small
\begin{adjustbox}{max width=\linewidth}
\begin{tabular}{lccccc}
\toprule
SN sample & $\alpha$ & $\dchi$ & $\sqrt{\dchi}$ & $P(\alpha<0)$ & $B_{\Lambda f}$ \\
\midrule
Pantheon+ & $0.085\pm0.060$ & $2.02$ & $1.42$ & $0.082$ & $3.17$ \\
DES-Y5 & $0.092\pm0.055$ & $2.86$ & $1.69$ & $0.049$ & $2.40$ \\
Union3 & $0.125\pm0.079$ & $2.40$ & $1.55$ & $0.057$ & $1.85$ \\
\bottomrule
\end{tabular}
\end{adjustbox}
\end{table}

\begin{table*}
\centering
\caption{Marginalised posteriors (mean $\pm1\sigma$) at $\eta=\sqrt6$. Each analysis combines CC + DESI DR2 BAO + Planck
distance priors with one SN compilation.}
\label{tab:post}
\begin{adjustbox}{max width=\linewidth}
\begin{tabular}{lcccc}
\toprule
Analysis & $\alpha$ & $\Om$ & $H_0$ [km\,s$^{-1}$\,Mpc$^{-1}$] & $\omega_b$ \\
\midrule
+ Pantheon+ & $0.085\pm0.060$ & $0.3068\pm0.0055$ & $67.65\pm0.60$ & $0.02258\pm0.00013$ \\
+ DES-Y5 & $0.092\pm0.055$ & $0.3073\pm0.0051$ & $67.59\pm0.55$ & $0.02258\pm0.00013$ \\
+ Union3 & $0.125\pm0.079$ & $0.3102\pm0.0071$ & $67.25\pm0.80$ & $0.02259\pm0.00013$ \\
\bottomrule
\end{tabular}
\end{adjustbox}
\end{table*}

\begin{table}
\centering
\caption{Goodness of fit and information criteria at $\eta=\sqrt6$, with $\Delta X\equiv X_\Lambda-X_f$ (positive
values favour the $f(Q)$ model) and $\Delta k=1$. $N$ is the total number of data points of each combination.}
\label{tab:ic}
\small
\begin{adjustbox}{max width=\linewidth}
\begin{tabular}{lcccccc}
\toprule
SN sample & $\chi^2_{\Lambda,\rm min}$ & $\chi^2_{f,\rm min}$ & $\dchi$ & $\Delta$AIC & $\Delta$BIC & $N$ \\
\midrule
Pantheon+ & 1433.87 & 1431.85 & $+2.02$ & $+0.02$ & $-5.38$ & 1638 \\
DES-Y5 & 1663.24 & 1660.38 & $+2.86$ & $+0.86$ & $-4.67$ & 1868 \\
Union3 & 56.49 & 54.09 & $+2.40$ & $+0.40$ & $-1.85$ & 70 \\
\bottomrule
\end{tabular}
\end{adjustbox}
\end{table}

\begin{table}
\centering
\caption{Contribution of each probe to $\dchi=\chi^2_\Lambda-\chi^2_f$ at the best fits, for $\eta=\sqrt6$ and for the
data-driven benchmark $\eta=1.6$ of Sec.~\ref{sec:eta}.}
\label{tab:perprobe}
\small
\begin{adjustbox}{max width=\linewidth}
\begin{tabular}{lccccc}
\toprule
Analysis & CC & BAO & CMB & SN & Total \\
\midrule
Pantheon+, $\sqrt6$ & $-0.10$ & $+1.09$ & $-1.83$ & $+2.85$ & $+2.02$ \\
DES-Y5, $\sqrt6$ & $-0.10$ & $+1.38$ & $-2.15$ & $+3.72$ & $+2.86$ \\
Union3, $\sqrt6$ & $-0.18$ & $+0.78$ & $-2.27$ & $+4.07$ & $+2.40$ \\
Pantheon+, $1.6$ & $-0.03$ & $+2.53$ & $-0.95$ & $+3.24$ & $+4.79$ \\
DES-Y5, $1.6$ & $-0.01$ & $+3.03$ & $-1.26$ & $+4.67$ & $+6.43$ \\
Union3, $1.6$ & $-0.06$ & $+2.25$ & $-0.53$ & $+5.29$ & $+6.94$ \\
\bottomrule
\end{tabular}
\end{adjustbox}
\end{table}

For all three SN compilations the best fit has $\alpha>0$: the improvement over \LCDM\ is $\dchi=2.02$, $2.86$ and
$2.40$, i.e.\ $1.4\sigma$, $1.7\sigma$ and $1.5\sigma$, and $92$--$95\%$ of the posterior lies at $\alpha>0$
(Table~\ref{tab:alpha}). The AIC is essentially neutral ($\Delta{\rm AIC}=+0.02$, $+0.86$, $+0.40$), while the BIC,
with its $\ln N$ penalty, favours \LCDM\ in all three cases ($\Delta{\rm BIC}=-5.38$, $-4.67$, $-1.85$;
Table~\ref{tab:ic}); the penalty is milder for Union3 because the binned sample gives $N=70$ instead of $1638$ or $1868$.
The Bayes factor, $B_{\Lambda f}=1.9$--$3.2$ ($|\ln B_{\Lambda f}|\leq1.2$), favours \LCDM, inconclusively or at most
weakly on the scale of \citet{Trotta2008}; through the Savage--Dickey ratio it scales with the width of the prior on
$\alpha$, here the full regularity range. At $\eta=\sqrt6$ the data are therefore compatible with $\alpha=0$. The best
fit has the same sign for the three compilations, but these share the CC, BAO and CMB data and part of the supernovae,
so they are a robustness check rather than three independent confirmations.

\begin{figure}
\centering
\includegraphics[width=\colfigwidth]{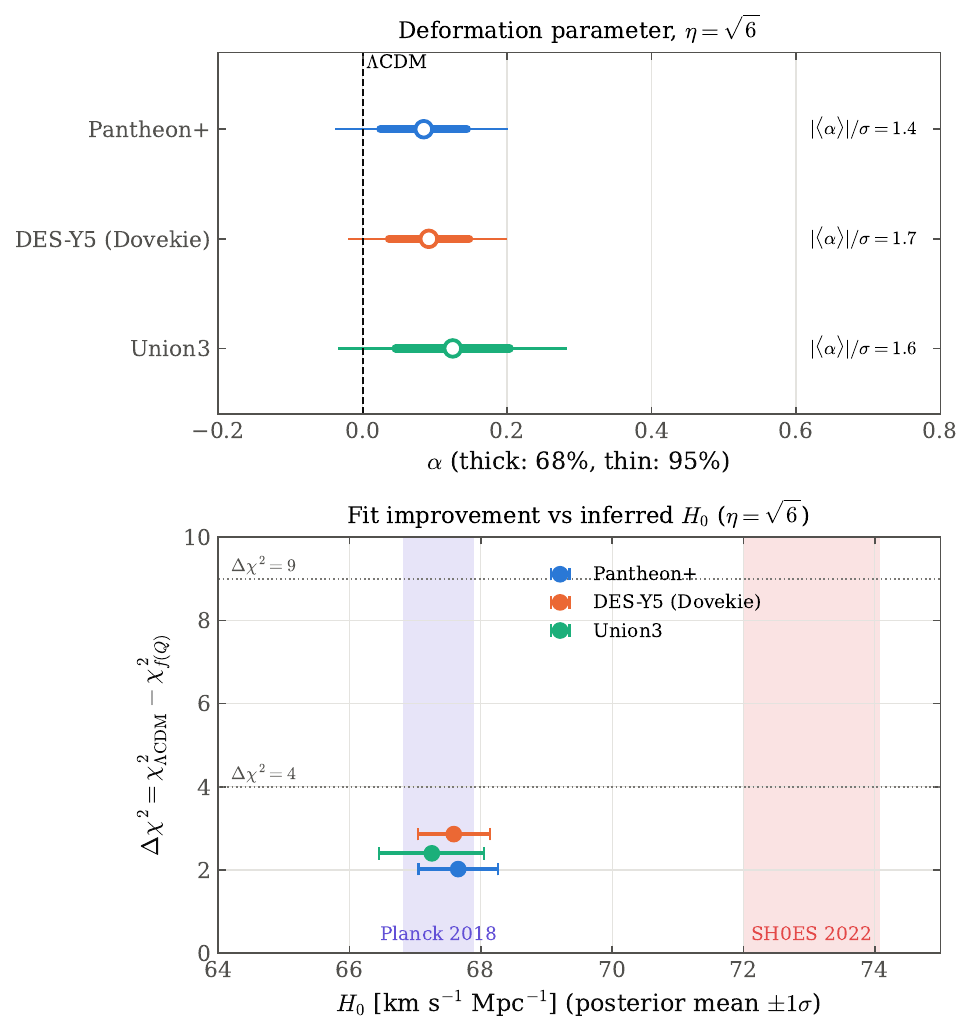}
\caption{Top: deformation parameter at $\eta=\sqrt6$ for the three SN compilations (thick: $68\%$, thin: $95\%$
intervals). Bottom: improvement of the fit over \LCDM\ against the inferred $H_0$, with the Planck 2018 and SH0ES 2022
bands.}
\label{fig:sncomp}
\end{figure}

\begin{figure}
\centering
\includegraphics[width=\colfigwidth]{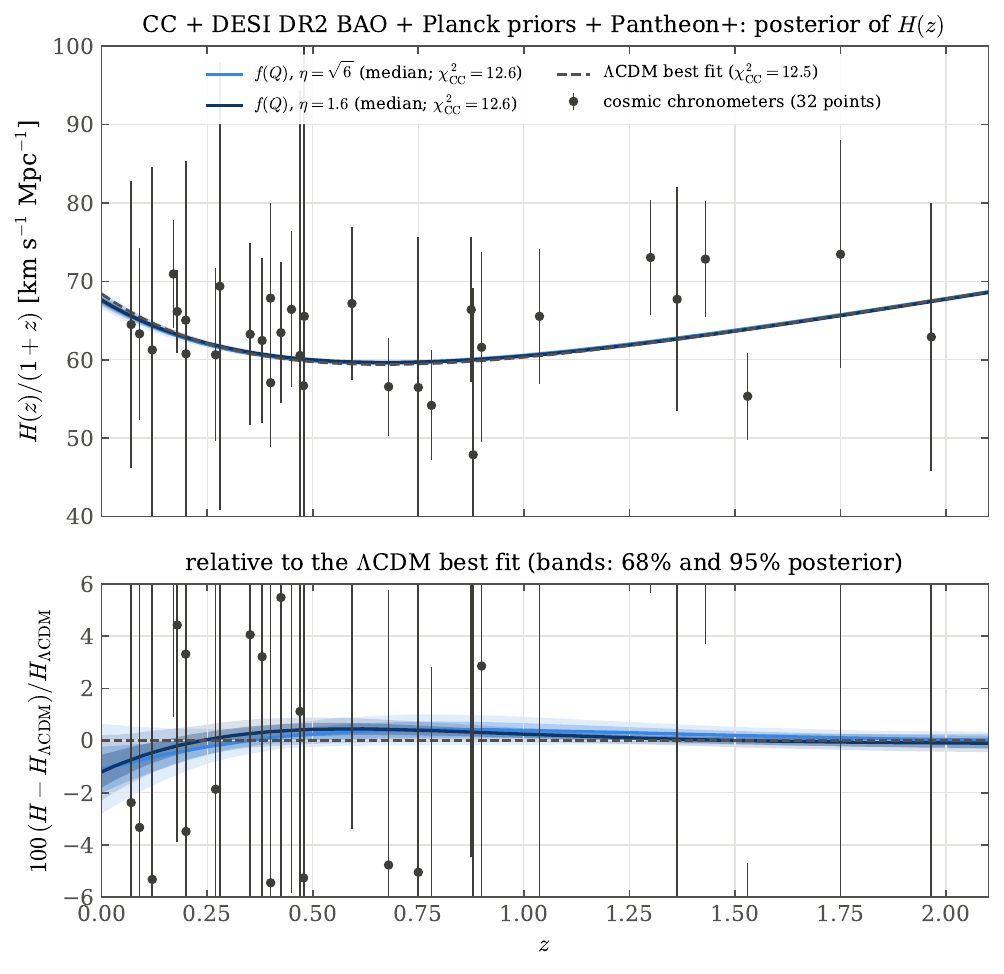}
\caption{Posterior of the expansion history for the Pantheon+ baseline at $\eta=\sqrt6$ and $\eta=1.6$, compared with
the 32 cosmic-chronometer measurements. Top: $H(z)/(1+z)$, with $68\%$ and $95\%$ bands and the \LCDM\ best fit.
Bottom: relative difference with respect to the \LCDM\ best fit; points and error bars beyond $\pm6\%$ are clipped.}
\label{fig:hz}
\end{figure}

The dependence on the SN compilation is weak: the mean value of $\alpha$ moves by about $0.5\sigma$ between Pantheon+ and
Union3. Table~\ref{tab:perprobe} shows where the small improvement comes from: the SN data gain
$\dchi\simeq3$--$4$ and the BAO data about $1$, while the CMB distance priors lose about $2$; the CC data are
essentially neutral. Figure~\ref{fig:sncomp} compares the three compilations: the inferred $H_0$ stays at
$67.3$--$67.7$\,km\,s$^{-1}$\,Mpc$^{-1}$, close to the Planck value~\citep{Planck2018VI} and far from the SH0ES
determination~\citep{Riess2022}, so the model does not address the Hubble tension~\citep{DiValentino2021}. Figure~\ref{fig:hz} compares the posterior expansion
history with the CC data: the deviation from \LCDM\ is below $1.2\%$ at all redshifts, much smaller than the CC
uncertainties ($\chi^2_{\rm CC}=12.6$ for the $f(Q)$ model and $12.5$ for \LCDM, for 32 points).

\section{Choice of $\eta$}
\label{sec:eta}

The scale $\mu$ is a free parameter of the model and is not fixed by the theory. Our reference value $\eta=\sqrt6$
corresponds to $\mu^2=Q_0$ and is fixed before any comparison with the data. Since $\alpha$ multiplies
$\mu^2=\eta^2H_0^2$, the observable effect of a given $\alpha$ increases with $\eta$: a unit change of $\alpha$
modifies $H(z)$ by up to $12\%$ at $\eta=\sqrt6$ and by $1.8\%$ at $\eta=1.6$, so the same data constrain $\alpha$ more
loosely at small $\eta$, and values of $\alpha$ are not directly comparable between different values of $\eta$.

\begin{figure}
\centering
\includegraphics[width=\colfigwidth]{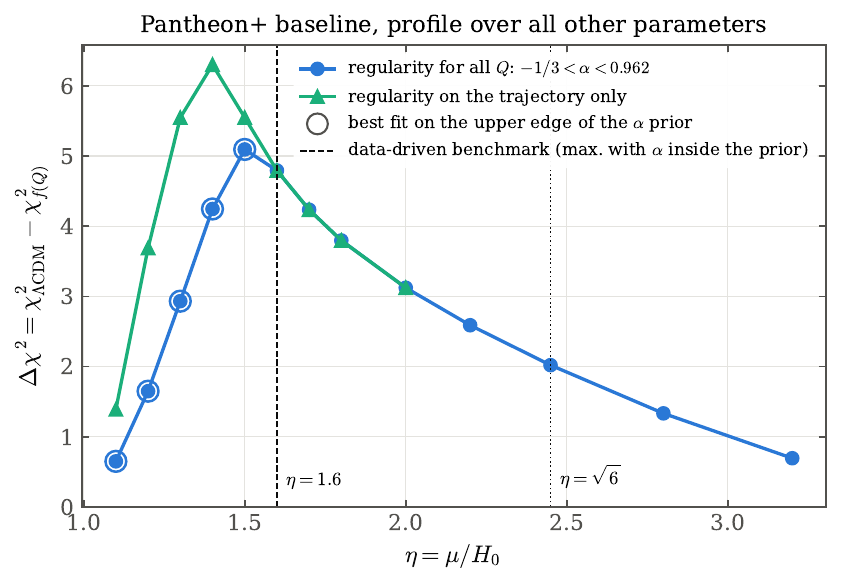}
\caption{Profile of $\dchi=\chi^2_\Lambda-\chi^2_f$ as a function of $\eta$ for the Pantheon+ baseline, minimised over all
other parameters, with regularity required for every $Q$ (blue) and only along the cosmological trajectory (green).
Open symbols mark best fits on the upper edge of the regularity range $\alpha<0.962$. The dashed line marks the data-driven benchmark
$\eta=1.6$, the dotted line the reference value $\eta=\sqrt6$.}
\label{fig:profile}
\end{figure}

\paragraph{Profile in $\eta$} To quantify how much the data can gain from adjusting $\mu$, we profiled the joint
$\chi^2$ over $\eta$ on a grid of step $0.1$, minimising over all other parameters with the regularity prior on
$\alpha$ (Fig.~\ref{fig:profile}). For the Pantheon+ baseline the improvement over \LCDM\ grows from $\dchi=2.02$ at
$\eta=\sqrt6$ to $4.79$ at $\eta=1.6$, where the best fit is $\alpha=0.83$. For $\eta\leq1.5$ the best fit lies on the
regularity bound $\alpha=0.962$, and the profile peaks there ($\dchi=5.09$ at $\eta=1.5$). The location of
this maximum is set by the regularity condition. If $D>0$ and $f_Q>0$ are required only along the cosmological
trajectory, from the radiation era to the future de Sitter point, rather than for every $Q$, larger values of $\alpha$
are allowed for $\eta\lesssim1.5$ and the maximum moves to $\eta=1.4$, with $\alpha=1.86$,
$\dchi=6.30$ and $z_{\rm cross}=0.56$ (Fig.~\ref{fig:profile}). The preferred value of $\eta$ therefore
depends on the regularity condition and not only on the data. We keep the stricter condition, which does not depend on
the cosmological trajectory.

\begin{figure*}
\centering
\includegraphics[width=0.80\textwidth]{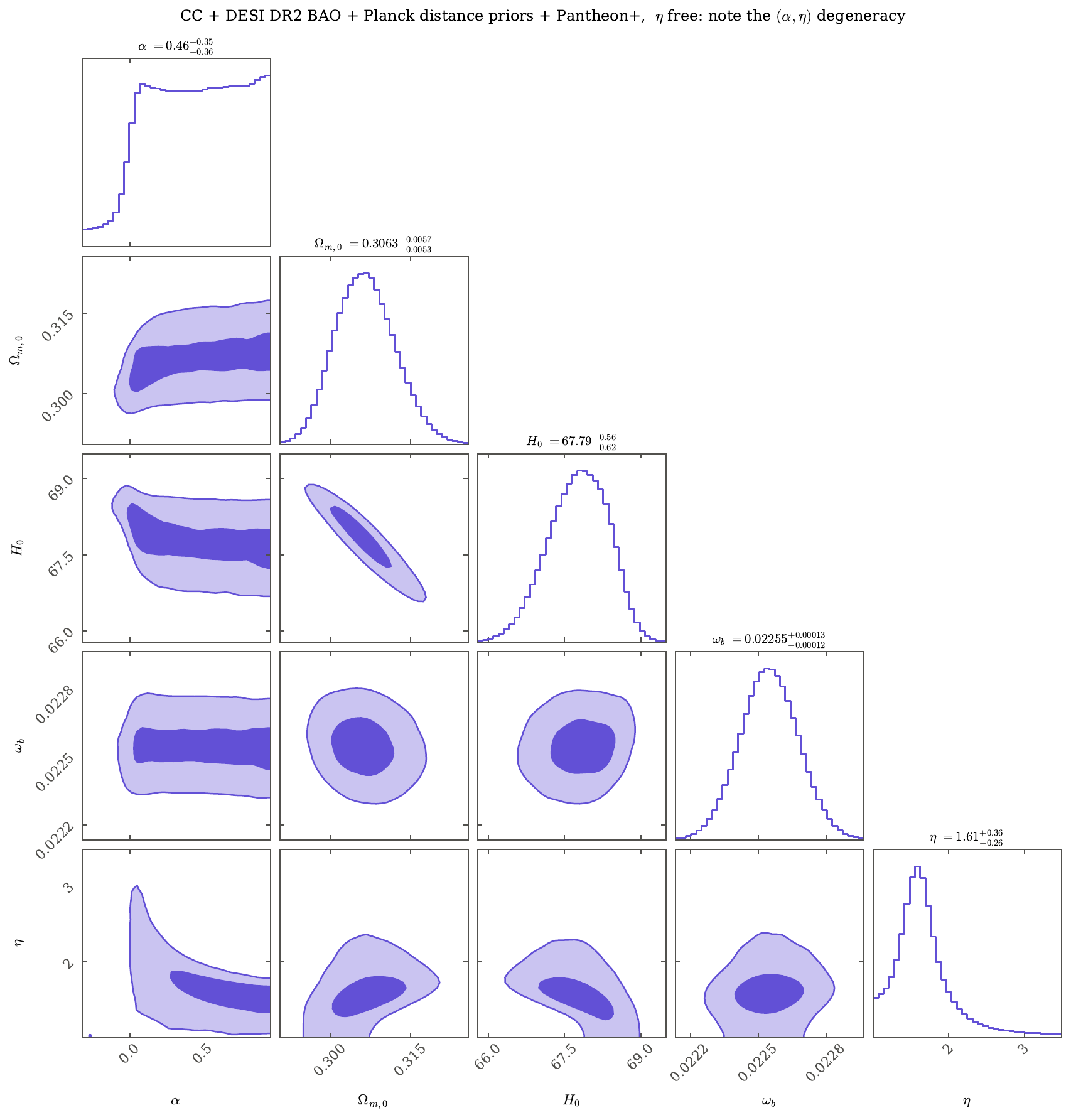}
\caption{Five-parameter posterior with $\eta$ free (Pantheon+ baseline, {\tt emcee}). The posterior of $\alpha$ is flat
between $0$ and the regularity bound, and the $(\alpha,\eta)$ degeneracy sets the location of the $\eta$ posterior.}
\label{fig:corner_free}
\end{figure*}

\begin{figure*}
\centering
\includegraphics[width=0.95\textwidth]{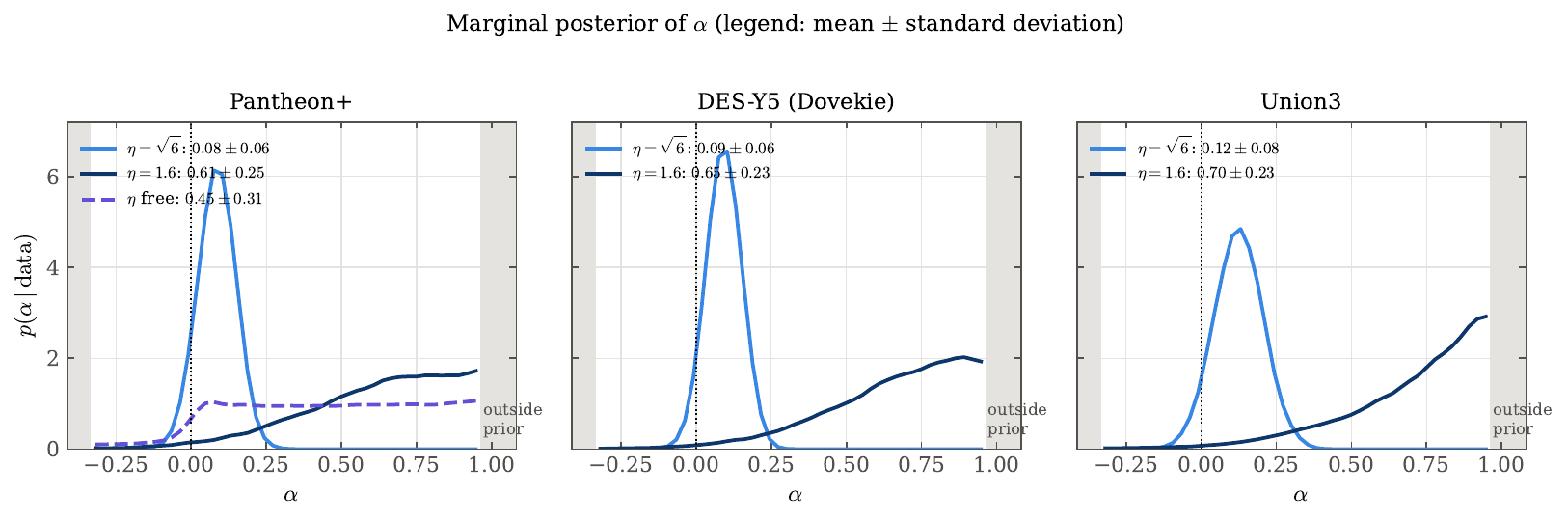}
\caption{Marginal posterior of $\alpha$ for the three SN compilations at $\eta=\sqrt6$ and $\eta=1.6$, and, for
Pantheon+, with $\eta$ free. The shaded regions lie outside the regularity range of Eq.~\eqref{eq:alpharange}.}
\label{fig:alpha1d}
\end{figure*}

\paragraph{$\eta$ as a free parameter} Leaving $\eta$ free in a five-parameter analysis (Fig.~\ref{fig:corner_free})
gives
\begin{equation}
\eta=1.61^{+0.36}_{-0.26}\,,\qquad P(\eta>\sqrt6)=0.05\,.
\end{equation}
The corresponding posterior of $\alpha$, $0.45\pm0.31$ (mean and standard deviation), is flat between $0$ and the
regularity bound (Fig.~\ref{fig:alpha1d}). Through the $(\alpha,\eta)$ degeneracy, the location of the $\eta$ posterior
is set by the upper bound of $\alpha$: it should not be read as a measurement of $\eta$. The Savage--Dickey ratio with
$\eta$ marginalised gives $B_{\Lambda f}=0.88$, which is inconclusive. The {\tt emcee} chain has a length of $61$
autocorrelation times of $\eta$ ($\tau_\eta\simeq653$ steps, effective sample size $\simeq1959$), and its
two halves give the same posterior mean of $\eta$ to $0.02$.

\begin{figure}
\centering
\includegraphics[width=\colfigwidth]{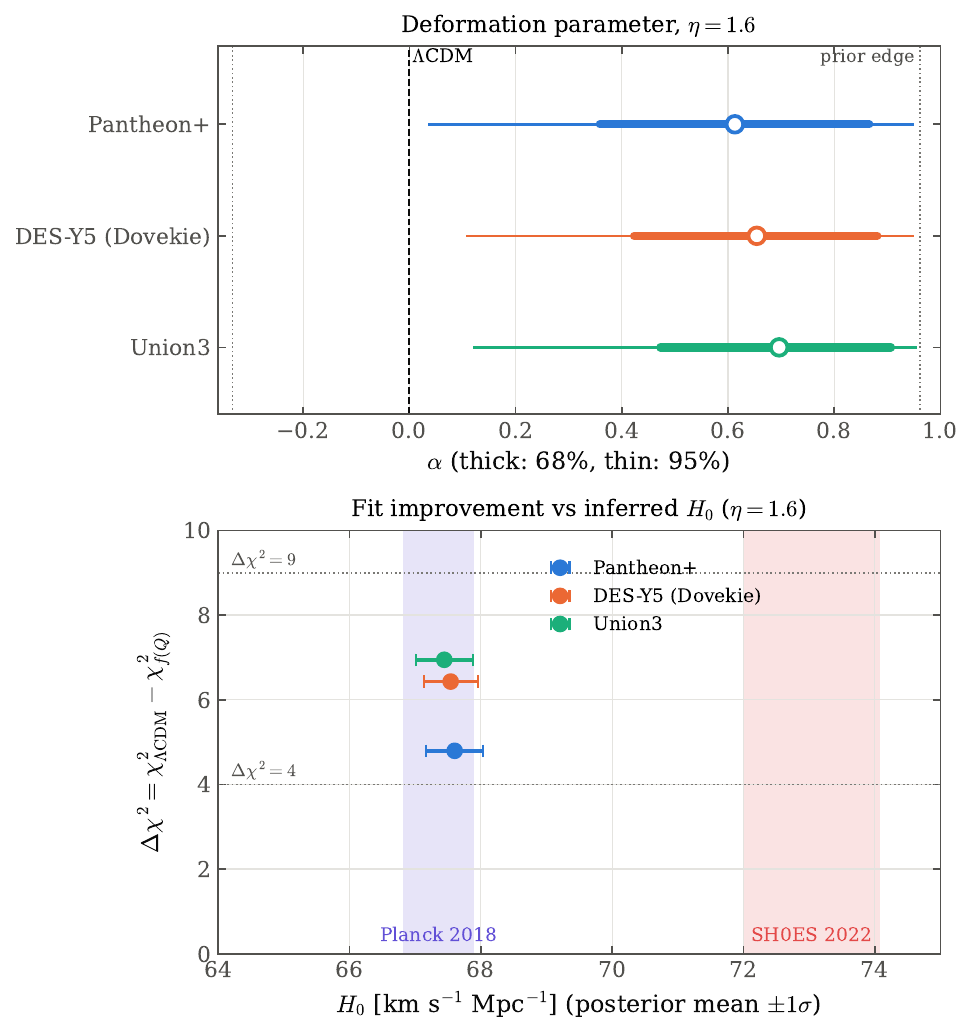}
\caption{As Fig.~\ref{fig:sncomp}, at $\eta=1.6$. The dotted lines mark the regularity range of $\alpha$.}
\label{fig:sncomp16}
\end{figure}

\paragraph{A data-driven benchmark} We adopt $\eta=1.6$, the maximum of the profile among the points whose best-fit
$\alpha$ lies inside the regularity range, as a second benchmark, and repeat the full analysis for it
(Table~\ref{tab:eta16} and Fig.~\ref{fig:sncomp16}; the full posteriors are shown in
\ref{app:eta16}). It is a selected benchmark, not the global maximum of the profile, which lies at the edge of the
regularity range. This value is also close to the median of the analysis with $\eta$ free. The crossing redshift moves
from $z_{\rm cross}=1.55$ at $\eta=\sqrt6$ to $0.78$ at $\eta=1.6$, closer to the low-redshift
crossing of the $w_0w_a$ reconstructions (Sec.~\ref{sec:models}).

Because this value is selected using the same data, $\eta$ is counted as an additional parameter in the information
criteria: $\Delta{\rm AIC}=+0.8$, $+2.4$, $+2.9$, positive for all three samples, and $\Delta{\rm BIC}=-10.0$, $-8.6$,
$-1.6$ (Table~\ref{tab:eta16}). At $\eta=1.6$ the improvement is $\dchi=4.79$, $6.43$ and $6.94$ (Pantheon+, DES-Y5,
Union3). Since $\eta$ is chosen from the data, the statistic to calibrate is the maximum of $\dchi$ over
$(\alpha,\eta)$, $5.09$, $6.49$ and $7.17$, reached at $\eta=1.5$, $1.5$ and $1.7$ with $\alpha$ on the regularity bound.
The Monte Carlo calibration of Sec.~\ref{sec:lee} gives $p=0.039$, $0.017$ and $0.012$, i.e.\ $2.1\sigma$, $2.4\sigma$
and $2.5\sigma$; a $\chi^2$ distribution with two degrees of freedom would give $1.8\sigma$, $2.1\sigma$ and $2.2\sigma$.
The $\chi^2$ rule is conservative here, because the templates at nearby $\eta$ are strongly correlated and the bound on
$\alpha$ limits the amplitude at small $\eta$ (\ref{app:lee}). At $\eta=1.6$ the posterior of $\alpha$ is broad,
$\alpha=0.61\pm0.25$, $0.65\pm0.23$ and $0.70\pm0.23$, and $9$--$15\%$ of it lies within $0.05$ of the regularity bound, so
these values depend on where the bound lies; for Union3 the best fit sits on the bound. The cosmological parameters are
almost unchanged with respect to $\eta=\sqrt6$ ($\Om=0.308\pm0.005$, $H_0=67.60\pm0.43$ for Pantheon+).

We therefore quote $\eta=\sqrt6$ as our main result and the $\eta=1.6$ benchmark as an indication of the largest
improvement the model can reach on current data. The two cases are not independent: they are the same data analysed
with two values of $\eta$.

\begin{table*}
\centering
\caption{Data-driven benchmark $\eta=1.6$. $\alpha$ is the marginalised mean and standard deviation, $\alpha_{\rm bf}$
the best fit, $\dchi$ the improvement at $\eta=1.6$. $\dchi_{\max}$ is the maximum over $(\alpha,\eta)$, reached at the
value of $\eta$ in brackets, and $n_\sigma^{\rm MC}$ its Monte Carlo significance (Sec.~\ref{sec:lee}; in brackets, the
value for a $\chi^2$ distribution with two degrees of freedom). The information criteria are given for $\Delta k=2$.}
\label{tab:eta16}
\begin{adjustbox}{max width=\linewidth}
\begin{tabular}{lcccccccc}
\toprule
SN sample & $\alpha$ & $\alpha_{\rm bf}$ & $\dchi$ & $\dchi_{\max}$ ($\eta$) & $n_\sigma^{\rm MC}$ & $\Delta$AIC & $\Delta$BIC & $z_{\rm cross}$ \\
\midrule
Pantheon+ & $0.61\pm0.25$ & $0.83$ & $4.79$ & $5.09$ ($1.5$) & $2.1$ ($1.8$) & $+0.8$ & $-10.0$ & $0.78$ \\
DES-Y5 & $0.65\pm0.23$ & $0.92$ & $6.43$ & $6.49$ ($1.5$) & $2.4$ ($2.1$) & $+2.4$ & $-8.6$ & $0.77$ \\
Union3 & $0.70\pm0.23$ & $0.962$ (bound) & $6.94$ & $7.17$ ($1.7$) & $2.5$ ($2.2$) & $+2.9$ & $-1.6$ & $0.77$ \\
\bottomrule
\end{tabular}
\end{adjustbox}
\end{table*}

\section{Growth of structure: an exploratory extension}
\label{sec:growth}

The growth of large-scale structure would provide a complementary handle on the deformation parameter. The
quasi-static equation used below, with $G_{\rm eff}=G/f_Q$, assumes a regime where perturbation theory is under control.
In the coincident-gauge branch used here perturbations around FLRW are infinitely strongly coupled~\citep{Gomes2024},
which does not guarantee that such a regime exists. We therefore treat Eq.~\eqref{eq:geff} as a phenomenological
prescription, conditional on a consistent completion of the theory, restrict ourselves to sub-horizon scales
($z\lesssim2$, $k\sim0.1\,h\,{\rm Mpc}^{-1}$), and keep the $f\sigma_8$ data out of the main analysis.

\subsection{Effective gravitational coupling}

In this regime the matter density contrast $\delta_m$ obeys~\citep{Khyllep2021,BeltranJimenez2020Cosmo,Heisenberg2024}
\begin{equation}
\delta_m''+\left(2+\frac{d\ln H}{dN}\right)\delta_m'=\frac32\,\Omega_m(a)\,\frac{G_{\rm eff}}{G}\,\delta_m\,,
\label{eq:growth}
\end{equation}
where primes denote derivatives with respect to $N=\ln a$, $\Omega_m(a)=\Om(1+z)^3/E^2(z)$, and
\begin{equation}
\frac{G_{\rm eff}}{G}=\frac{1}{f_Q}=\left[1+\alpha\,\frac{3-x}{(1+x)^3}\right]^{-1}.
\label{eq:geff}
\end{equation}
At early times ($x\to\infty$) the correction falls off as $\alpha/x^2$ and standard clustering is recovered. At the present
epoch with $\eta=\sqrt6$ ($x_0=1$), $G_{\rm eff}/G=1/(1+\alpha/4)$: for $\alpha=0.086$ the effective coupling is
suppressed by about $2\%$. The regularity range of Eq.~\eqref{eq:alpharange} guarantees $f_Q>0$, so $G_{\rm eff}$ is finite
and positive along the whole evolution.

We integrate Eq.~\eqref{eq:growth} numerically from $z=1000$, starting from the growing mode of the M\'esz\'aros equation
with radiation, and compute
\begin{equation}
f\sigma_8(z)=\sigma_{8,0}\,f(z)\,\frac{\delta_m(z)}{\delta_m(0)}\,,\qquad f\equiv\frac{d\ln\delta_m}{d\ln a}\,,
\end{equation}
with $\sigma_{8,0}$ a free parameter. This full solution is used in all the fits that include growth data. Adding the
Gold-2017 data to the main analysis at $\eta=\sqrt6$ gives $\alpha=0.087\pm0.060$ and $\sigma_{8,0}=0.756\pm0.029$ with
Pantheon+; for the three SN compilations and both values of $\eta$ it lowers $\dchi$ by $0.06$--$0.3$ and moves the
best-fit $\alpha$ by less than $0.03$. At the present precision the growth data carry no information on the
deformation.

\subsection{Growth-index approximation}

\begin{figure*}
\centering
\includegraphics[width=0.92\textwidth]{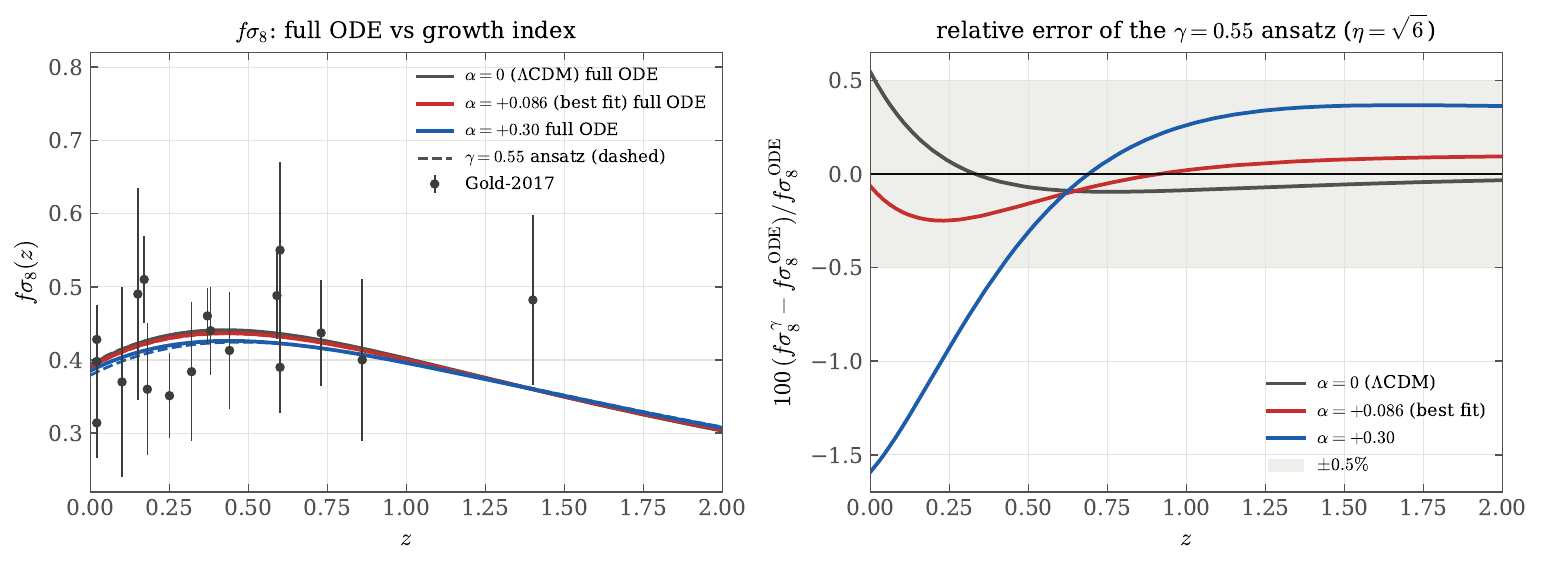}
\caption{Growth-index approximation against the full solution of Eq.~\eqref{eq:growth} at $\eta=\sqrt6$. Left:
$f\sigma_8(z)$ from the full equation (solid) and from the $\gamma=0.55$ ansatz (dashed) for $\alpha=0$, for the best fit
$\alpha=0.086$ and for $\alpha=0.30$, with the Gold-2017 data. Right: relative difference between the two, in per cent.}
\label{fig:growthval}
\end{figure*}

A common shortcut is the growth-index parametrization~\citep{Linder2005,LinderCahn2007,Tsujikawa2009}
\begin{equation}
f(a)\simeq\left[\Omega_m(a)\,\frac{G_{\rm eff}(a)}{G}\right]^\gamma,\qquad\gamma=0.55\,,
\label{eq:gamma}
\end{equation}
originally derived for \LCDM\ and mild modifications of gravity. Figure~\ref{fig:growthval} compares it with the full
solution of Eq.~\eqref{eq:growth}. At the best fit $\alpha=0.086$ the approximation reproduces $f\sigma_8$ to better than
$0.3\%$ over $0<z<2$; the largest error, $0.5\%$ at $z=0$, is the one already present in \LCDM, and for the extreme value
$\alpha=0.30$ the error reaches $1.6\%$ at $z=0$. These errors are well below the uncertainties of the RSD data, so the
approximation would be adequate; we nevertheless use the full equation.

\subsection{Local constraints and gravitational waves}

If the cosmological $G_{\rm eff}$ of Eq.~\eqref{eq:geff} also controlled gravity in the Solar System, its time variation
would be, at $\eta=\sqrt6$,
\begin{equation}
\frac{\dot G_{\rm eff}}{G_{\rm eff}}=-H\,\frac{d\ln f_Q}{dx}\,\frac{dx}{dN}\simeq-0.47\,\alpha\,H_0\,,
\end{equation}
i.e.\ $|\dot G/G|\simeq3\times10^{-12}\,{\rm yr}^{-1}$ for $\alpha=0.086$, about forty times the uncertainty of the Lunar
Laser Ranging (LLR) constraint $\dot G/G=(7.1\pm7.6)\times10^{-14}\,{\rm yr}^{-1}$~\citep{Hofmann2018}, and about
three hundred times that of the most recent LLR analysis, $\dot G/G=(-5.0\pm9.6)\times10^{-15}\,{\rm yr}^{-1}$~\citep{Biskupek2021}. A viable embedding of the model
would therefore require a mechanism that decouples the local value of the gravitational coupling from the cosmological
one. No such screening mechanism has been shown to exist in $f(Q)$ gravity, so the model should be read as a
description of the expansion history, not as a theory tested in the Solar System.

Tensor modes propagate at the speed of light in $f(Q)$ gravity~\citep{BeltranJimenez2020Cosmo}, in agreement with
GW170817~\citep{LIGO2017,LIGO2017GRB,Ezquiaga2017}, but their friction term is modified by
$f_Q$~\citep{BeltranJimenez2020Cosmo,Frusciante2021}, so that the gravitational-wave and electromagnetic luminosity
distances differ:
\begin{equation}
\frac{d_L^{\rm GW}(z)}{d_L^{\rm EM}(z)}=\sqrt{\frac{f_Q(z=0)}{f_Q(z)}}\,.
\end{equation}
For $\alpha=0.086$ at $\eta=\sqrt6$ the ratio approaches $\sqrt{1+\alpha/4}\simeq1.01$ at high redshift, below the
precision of current standard sirens but within reach of future detectors.

\section{Comparison with other models}
\label{sec:models}

The discussion of evolving dark energy prompted by DESI is often expressed in terms of the Chevallier--Polarski--Linder
(CPL) parametrization~\citep{Chevallier2001,Linder2003},
\begin{equation}
w(a)=w_0+w_a(1-a)=w_0+w_a\,\frac{z}{1+z}\,,
\label{eq:cpl}
\end{equation}
and of model-independent reconstructions~\citep{Calderon2024,Lodha2025DR1,Lodha2025DR2}. CPL provides a flexible
description of the equation of state but is not tied to a Lagrangian; the model considered here, by contrast, produces a
qualitatively similar evolution from a one-parameter deformation of the gravitational action. To compare the two on an
equal footing, and to place the model among other simple extensions of \LCDM, we fitted the following models to exactly
the same data and with the same pipeline:
$w$CDM (constant $w$); $w_0w_a$CDM [Eq.~\eqref{eq:cpl}, with the DESI priors~\citep{DESI2025DR2} $w_0\in(-3,1)$,
$w_a\in(-3,2)$, $w_0+w_a<0$]; the power-law $f(Q)=Q+M\,Q_0(Q/Q_0)^n$~\citep{BeltranJimenez2020Cosmo,Khyllep2021}, for which
$E^2=\Om(1+z)^3+\Or(1+z)^4+(1-\Om-\Or)E^{2n}$ and $n=0$ is \LCDM; and the exponential
$f(Q)=Q\,e^{\lambda Q_0/Q}$~\citep{Anagnostopoulos2021}, for which
$(E^2-2\lambda)\,e^{\lambda/E^2}=\Om(1+z)^3+\Or(1+z)^4$, with $\lambda$ fixed by $E(0)=1$, so that the model has the same
number of parameters as \LCDM. All models are fitted to the data of the main analysis, without growth data.

\begin{table*}
\centering
\caption{Comparison with other models on the same data and pipeline. $\Delta X=X_\Lambda-X_{\rm model}$ (positive values
favour the model), given as Pantheon+ / DES-Y5 / Union3; $\Delta k$ is the number of additional parameters with respect to
\LCDM. For $\eta=1.6$, $\eta$ is counted as a parameter.}
\label{tab:models}
\begin{adjustbox}{max width=\linewidth}
\begin{tabular}{lcccc}
\toprule
Model & $\Delta k$ & $\dchi$ & $\Delta$AIC & $\Delta$BIC \\
\midrule
$f(Q)$ Hu--Sawicki-inspired, $\eta=\sqrt6$ & 1 & $+2.0$ / $+2.9$ / $+2.4$ & $0.0$ / $+0.9$ / $+0.4$ & $-5.4$ / $-4.7$ / $-1.8$ \\
$f(Q)$ Hu--Sawicki-inspired, $\eta=1.6$ & 2 & $+4.8$ / $+6.4$ / $+6.9$ & $+0.8$ / $+2.4$ / $+2.9$ & $-10.0$ / $-8.6$ / $-1.6$ \\
$w_0w_a$CDM (CPL) & 2 & $+7.9$ / $+11.2$ / $+14.1$ & $+3.9$ / $+7.2$ / $+10.1$ & $-6.9$ / $-3.9$ / $+5.6$ \\
$w$CDM & 1 & $+0.8$ / $+1.6$ / $+0.8$ & $-1.2$ / $-0.4$ / $-1.2$ & $-6.6$ / $-6.0$ / $-3.5$ \\
$f(Q)=Q+MQ_0(Q/Q_0)^n$ & 1 & $+0.1$ / $+0.3$ / $0.0$ & $-1.9$ / $-1.7$ / $-2.0$ & $-7.3$ / $-7.2$ / $-4.2$ \\
$f(Q)=Q\,e^{\lambda Q_0/Q}$ & 0 & $-51.6$ / $-63.8$ / $-37.3$ & $-51.6$ / $-63.8$ / $-37.3$ & $-51.6$ / $-63.8$ / $-37.3$ \\
\bottomrule
\end{tabular}
\end{adjustbox}
\end{table*}

The results are summarised in Table~\ref{tab:models}. Among the one-parameter extensions, the Hu--Sawicki-inspired model
at $\eta=\sqrt6$ gives the largest improvement: $w$CDM gains $\dchi=0.8$--$1.6$, the power-law $f(Q)$ model converges to
$n=0.01$--$0.02$, i.e.\ to \LCDM, and the exponential $f(Q)$ model is strongly disfavoured, mainly by the SN and BAO
data. The $w_0w_a$ parametrization, with one parameter more, improves the fit by $\dchi=7.9$--$14.1$, with best fits
$(w_0,w_a)=(-0.85,-0.53)$ with Pantheon+, $(-0.82,-0.65)$ with DES-Y5 and $(-0.68,-1.00)$ with Union3, close to the DESI
DR2 results~\citep{DESI2025DR2}. These improvements are smaller than those reported by DESI with the full Planck
likelihood, which correspond to $2.8$--$4.2\sigma$~\citep{DESI2025DR2}, because the CMB enters here only through the
compressed priors and because the Dovekie recalibration reduces the preference of the DES SNe~\citep{Popovic2026}. With a
single Lagrangian parameter, our model recovers between a sixth (Union3) and a quarter (Pantheon+, DES-Y5) of the $w_0w_a$
improvement at $\eta=\sqrt6$ and about half at $\eta=1.6$, where, counting $\eta$, it has as many parameters as $w_0w_a$, whose $\Delta$AIC remains higher.

\begin{figure*}
\centering
\includegraphics[width=0.92\textwidth]{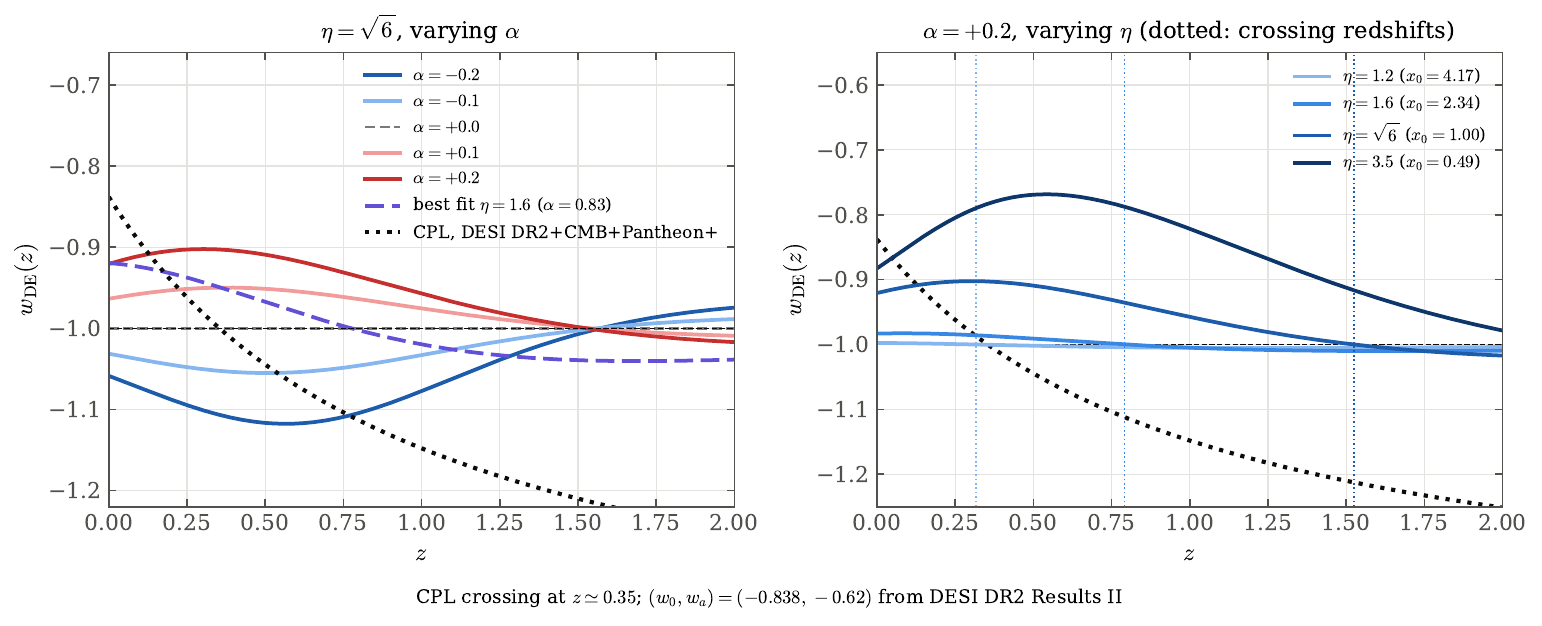}
\caption{Effective equation of state of the model compared with the CPL best fit of DESI DR2+CMB+Pantheon+,
$(w_0,w_a)=(-0.838,-0.62)$~\citep{DESI2025DR2} (dotted). Left: $\eta=\sqrt6$ with varying $\alpha$, and the best fit at
$\eta=1.6$ (dashed). Right: $\alpha=0.2$ with varying $\eta$ inside the prior range; the vertical lines mark the crossing
redshifts. The crossing redshift is controlled by $\eta$ rather than by $\alpha$.}
\label{fig:cpl}
\end{figure*}

The reason is visible in Fig.~\ref{fig:cpl}: the data prefer a phantom crossing at low redshift, as in the CPL best fit,
which crosses $w=-1$ at $z\simeq0.35$, whereas the model crosses at $z_{\rm cross}\simeq1.55$ for $\eta=\sqrt6$ and
$0.78$ for $\eta=1.6$. In addition, the CPL fits improve the agreement with the CMB distance priors
($\dchi_{\rm CMB}\simeq+2.4$ with Pantheon+), while the $f(Q)$ deformation slightly worsens it
(Table~\ref{tab:perprobe}). The crossing redshift of the model is fixed by $\eta$ through $x_+$; a crossing at $z\simeq0.35$
would require $\eta\simeq1.2$, where the effect of $\alpha$ on the expansion history is weak and the best fit sits on the
regularity bound, with $\dchi=1.65$ only; with regularity required only along the trajectory the improvement
at $\eta=1.2$ reaches $\dchi=3.69$ (Fig.~\ref{fig:profile}).

\section{Sampler robustness and numerical cross-checks}
\label{sec:robust}

\begin{figure*}
\centering
\includegraphics[width=0.80\textwidth]{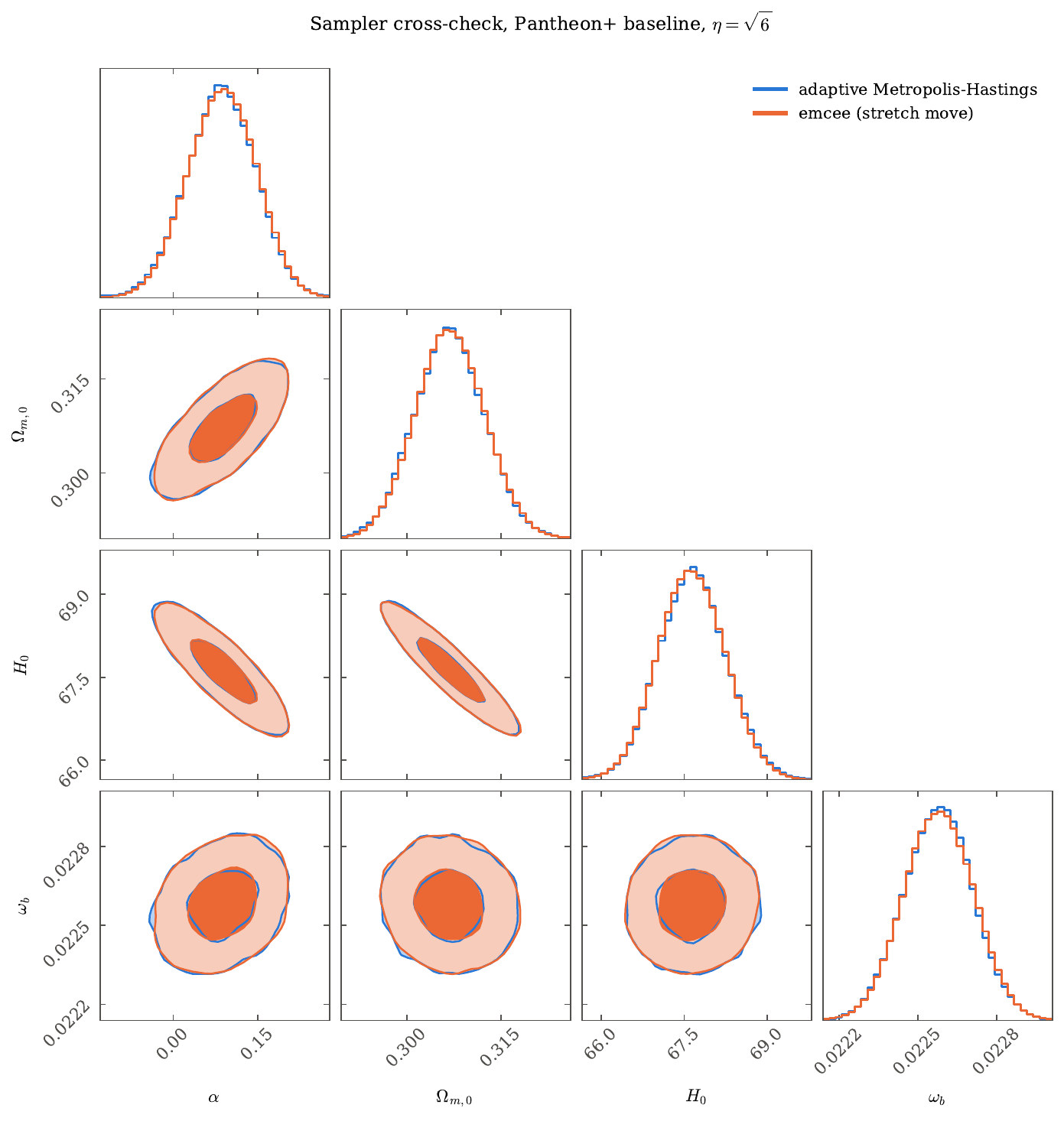}
\caption{Sampler cross-check for the Pantheon+ baseline at $\eta=\sqrt6$: adaptive Metropolis--Hastings (blue) and
{\tt emcee} (orange).}
\label{fig:sampler}
\end{figure*}

A preference inferred from an MCMC analysis must be checked against possible dependence on the sampling algorithm. We
repeated the baseline four-parameter analysis at $\eta=\sqrt6$ with {\tt emcee}~\citep{Goodman2010,ForemanMackey2013}
(32 walkers, $1000$ burn-in and $4000$ production steps, i.e.\ more than $80$ autocorrelation times). The
result,
\begin{equation}
\alpha=0.086\pm0.059\,,
\end{equation}
agrees with the adaptive Metropolis--Hastings baseline, $0.085\pm0.060$, within $0.03\sigma$, and the one- and
two-dimensional posteriors of the two samplers coincide (Fig.~\ref{fig:sampler}).

We also checked the impact of the analysis choices. Using a different treatment of some likelihood ingredients
(12 instead of 13 DESI points, the 1580 Pantheon+ SNe without the calibrators and $D_L=(1+z_{\rm HD})D_M$) changes
$\dchi$ by at most $0.25$ at $\eta=\sqrt6$ and $0.45$ at $\eta=1.6$; computing $z_*$ at each step instead of fixing it
and adding the growth data change $\dchi$ by at most $0.05$ and $0.3$, respectively (\ref{app:robust}). The CMB block
was validated against CAMB in the \LCDM\ limit (\ref{app:cmb}). Finally, the best fits and the profile in $\eta$ were obtained with several
starting points, and the numerical solution of Eq.~\eqref{eq:constraint} was verified to satisfy the constraint to a
relative accuracy of $10^{-9}$ on the branch that passes through $x_0$ today.

\section{Discussion}
\label{sec:discussion}

Three main points emerge from the analysis.

First, the model provides a transparent realization of phantom crossing. As for any deformation linear in $\alpha$,
the crossing points, expressed in terms of $x$, do not depend on the deformation amplitude; what the specific form of
Eq.~\eqref{eq:model} adds is a single crossing on the past trajectory, in the direction preferred by the DESI $w_0w_a$
fits.
Which root is reached, and at which redshift, depends on the scale $\eta$: for $1.01<\eta<5.9$ the past trajectory
crosses only $x_+$, at a redshift that increases with $\eta$. In this sense the mechanism differs from phenomenological
parametrizations such as CPL, where the crossing redshift is a fitted quantity, and from effective-field or kinetic
braiding realizations of evolving dark energy, where the crossing is generated by additional scalar
dynamics~\citep{Deffayet2010,Cai2010}.

Second, current data are compatible with \LCDM. At the reference scale $\eta=\sqrt6$ the improvement over \LCDM\ is
$\dchi=2.0$--$2.9$, i.e.\ $1.4$--$1.7\sigma$, with $\alpha>0$ for all three SN compilations; the AIC is neutral, and the
Bayes factor and the BIC favour \LCDM. Allowing the data to choose $\eta$ increases the improvement to
$\dchi=4.8$--$6.9$ at $\eta=1.6$; accounting for the search over $\eta$, Monte Carlo simulations give
$2.1$--$2.5\sigma$, and the preferred values of $\alpha$ approach the regularity bound, so that the preferred $\eta$
depends on the regularity condition. On the same data the model performs better than $w$CDM and than the power-law and
exponential $f(Q)$ models, and recovers part of the improvement of the $w_0w_a$ parametrization from a Lagrangian; the
remaining difference reflects its higher crossing redshift. The weak dependence on the SN compilation, with $\alpha$
moving by about $0.5\sigma$, suggests that the deviation is driven by the combination of SN, BAO and CMB information
rather than by any single probe; the three compilations, however, share all the other data and are not independent
confirmations.

Third, several caveats limit the interpretation. The most important is theoretical: \citet{Gomes2024} have
shown that perturbations around FLRW in nonlinear $f(Q)$ theories are infinitely strongly coupled in the coincident-gauge
branch used here and in one of the other two flat branches, while the remaining branch propagates a ghost. An infinitely strongly coupled
background does not guarantee a range of scales where perturbation theory holds, so the quasi-static expression
$G_{\rm eff}=G/f_Q$ is not a controlled prediction of the theory. Our main results therefore concern the background, and
the growth constraints of Sec.~\ref{sec:growth} are conditional on a consistent completion of the theory, which a
covariant reformulation does not provide automatically. Second, if the cosmological $G_{\rm eff}$ applied locally, Lunar
Laser Ranging would exclude the values of $\alpha$ considered here; since no screening mechanism is known for $f(Q)$
gravity, the model is at present a description of the expansion history rather than a theory tested in the Solar System.
Third, the CMB enters only through compressed distance priors, which do not capture the late-time integrated
Sachs--Wolfe effect or CMB lensing; a full Boltzmann analysis could tighten or shift the constraints. Finally, the sign
of the deviation is conditional on the current supernova compilations and may change as their systematics are further
refined.

\section{Conclusions}
\label{sec:conclusions}

We have introduced and tested a one-parameter extension of the symmetric teleparallel equivalent of general relativity,
\begin{equation}
f(Q)=Q+2\Lambda+\alpha\mu^2\,\frac{x-1}{(1+x)^2}\,,\qquad x=\frac{Q}{\mu^2}=\frac{6H^2}{\mu^2}\,,
\end{equation}
inspired by the Hu--Sawicki class of $f(R)$ models~\citep{HuSawicki2007}. Our main results are the following.
\begin{enumerate}
\item \emph{Algebraic phantom crossing.} The effective equation of state crosses $w_{\rm DE}=-1$ at
$x_\pm=3\pm2\sqrt2$, independent of $\alpha$ and $\mu$ as for any deformation linear in $\alpha$. For
$1.01<\eta<5.9$ only $x_+$ is reached by the past trajectory, in the direction found in the DESI fits; the crossing
redshift is set by $\eta$, with $z_{\rm cross}=1.55$ at $\eta=\sqrt6$ and $0.78$ at $\eta=1.6$.
\item \emph{Regularity.} Requiring $f_Q>0$ and an invertible Friedmann constraint for every $Q$ restricts the
deformation to $-1/3<\alpha<0.962$, independently of $\eta$; a future de Sitter point requires in addition
Eq.~\eqref{eq:desitter}.
\item \emph{No significant deviation at $\eta=\sqrt6$.} The baseline combination CC + DESI DR2 BAO + Planck
distance priors + Pantheon+ gives $\alpha=0.085\pm0.060$, with $92\%$ of the posterior at $\alpha>0$, and
$\dchi=2.02$ with respect to \LCDM, i.e.\ $1.4\sigma$; $\Delta{\rm AIC}=+0.02$, $\Delta{\rm BIC}=-5.38$ and
$B_{\Lambda f}=3.2$ in favour of \LCDM. With DES-Y5 and Union3 we find $1.7\sigma$ and $1.5\sigma$, with the same sign
of the deviation. The {\tt emcee} cross-check gives $\alpha=0.086\pm0.059$.
\item \emph{Data-driven scale.} Choosing $\eta$ from the profile of the data gives $\eta=1.6$, with $\dchi=4.8$, $6.4$
and $6.9$ for the three SN compilations; accounting for the search over $\eta$ with Monte Carlo simulations, the
maximum improvement corresponds to $2.1$--$2.5\sigma$. The analysis with $\eta$ free gives $\eta=1.61^{+0.36}_{-0.26}$, with
$\alpha$ limited by the regularity bound, and the preferred $\eta$ depends on the regularity condition.
\item \emph{Comparison with other models.} On the same data the model gives the largest improvement among the one-parameter
extensions considered ($w$CDM, power-law and exponential $f(Q)$) and recovers between a sixth and a half of the improvement of
$w_0w_a$, whose crossing occurs at $z\simeq0.35$.
\item \emph{Background close to \LCDM.} The expansion history differs from \LCDM\ by less than $1.5\%$, and the inferred
$H_0=67.3$--$67.7$\,km\,s$^{-1}$\,Mpc$^{-1}$ does not move towards the SH0ES value: the model does not address the Hubble
tension.
\end{enumerate}

The model thus provides a compact, geometrical realization of an evolving dark-energy equation of state with a
structural phantom crossing, controlled by a single dimensionless parameter once the scale $\eta$ is fixed. Current data are
compatible with \LCDM: the deviation is at the $1.4$--$1.7\sigma$ level for $\eta$ fixed a priori and $2.1$--$2.5\sigma$
when $\eta$ is chosen by the data, which forthcoming DESI and supernova data will be able to confirm or rule out. The
viability of the model beyond the background level depends on the resolution of the strong-coupling problem of $f(Q)$
cosmologies and on the existence of a mechanism that protects local tests of gravity, which has not been shown; both are
natural directions for future work, together with a full Boltzmann treatment.

\section*{CRediT authorship contribution statement}
\textcolor{black}{\ Giovanni Tuccio: Conceptualization, Methodology, Software, Formal analysis,
Writing -- original draft. Giovanni Montani: Conceptualization, Supervision, Writing -- review \& editing.
Alessandro Melchiorri: Supervision, Writing -- review \& editing.}

\section*{Acknowledgments}
During the preparation of this work, the
authors used the AI technologies Claude 
to polish the manuscript. All scientific ideas, analysis and
interpretation of the results are the work of the authors,
who take full responsibility for the final text.

\section*{Declaration of competing interest}
The authors declare that they have no known competing financial interests or personal relationships that could have
appeared to influence the work reported in this paper.

\section*{Data and code availability}
All data sets used in this work are public. Code available upon request.

\appendix
\section{Posteriors at $\eta=1.6$}
\label{app:eta16}

\begin{figure*}
\centering
\includegraphics[width=0.80\textwidth]{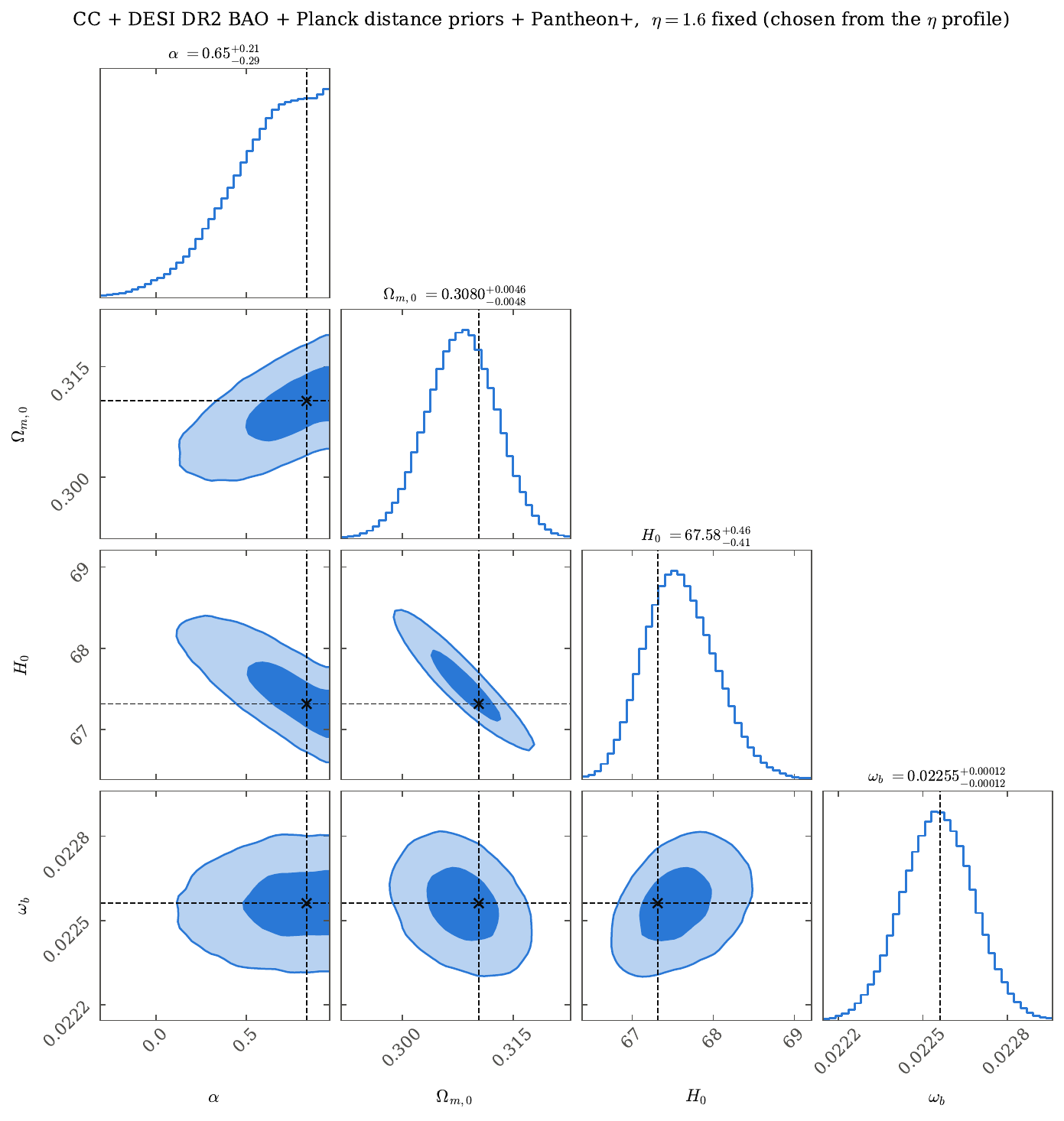}
\caption{As Fig.~\ref{fig:corner_p}, at the data-driven benchmark $\eta=1.6$. The posterior of $\alpha$ is cut by the
regularity bound $\alpha<0.962$.}
\label{fig:corner_p16}
\end{figure*}

\begin{figure*}
\centering
\includegraphics[width=0.49\textwidth]{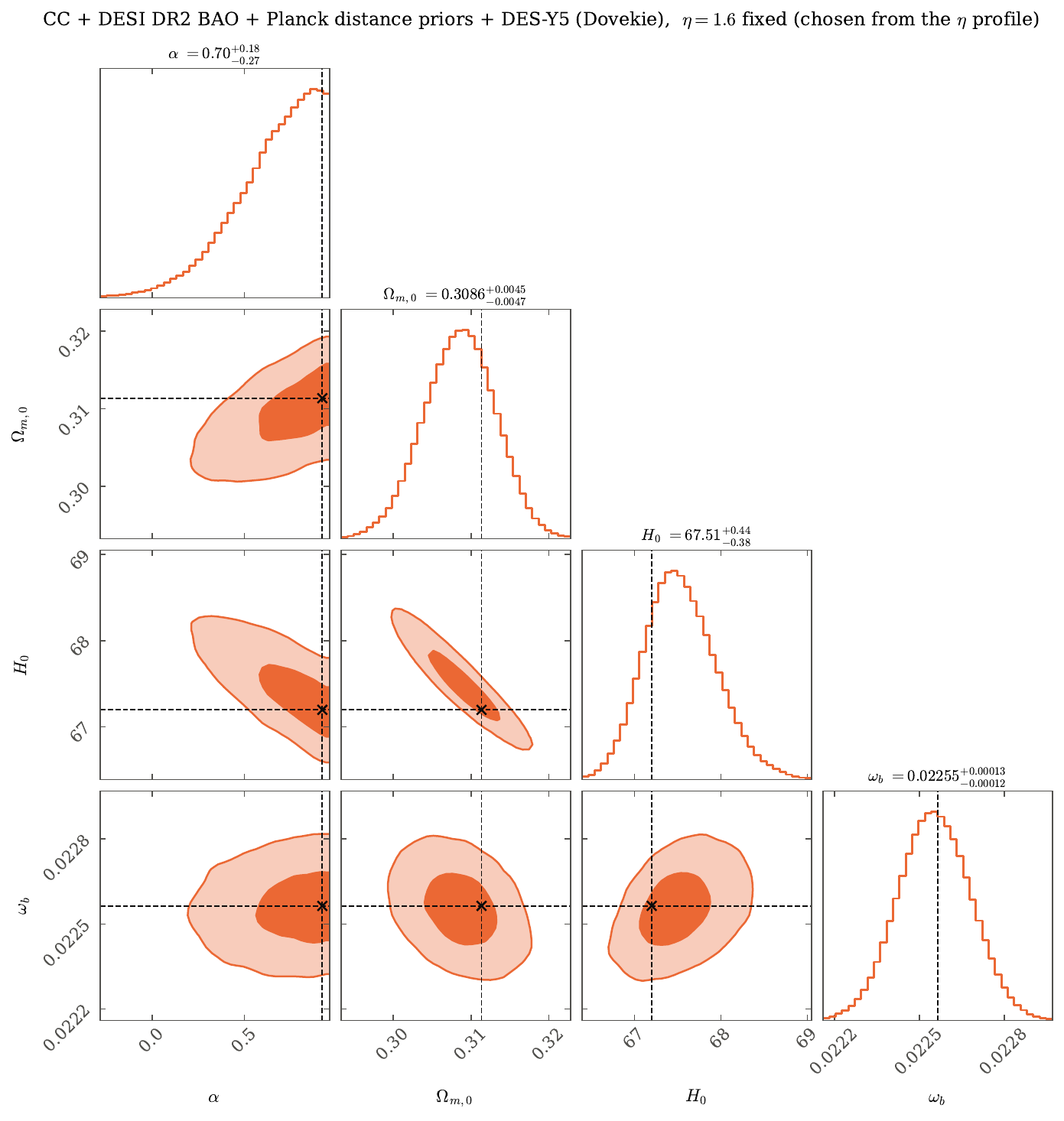}\hfill
\includegraphics[width=0.49\textwidth]{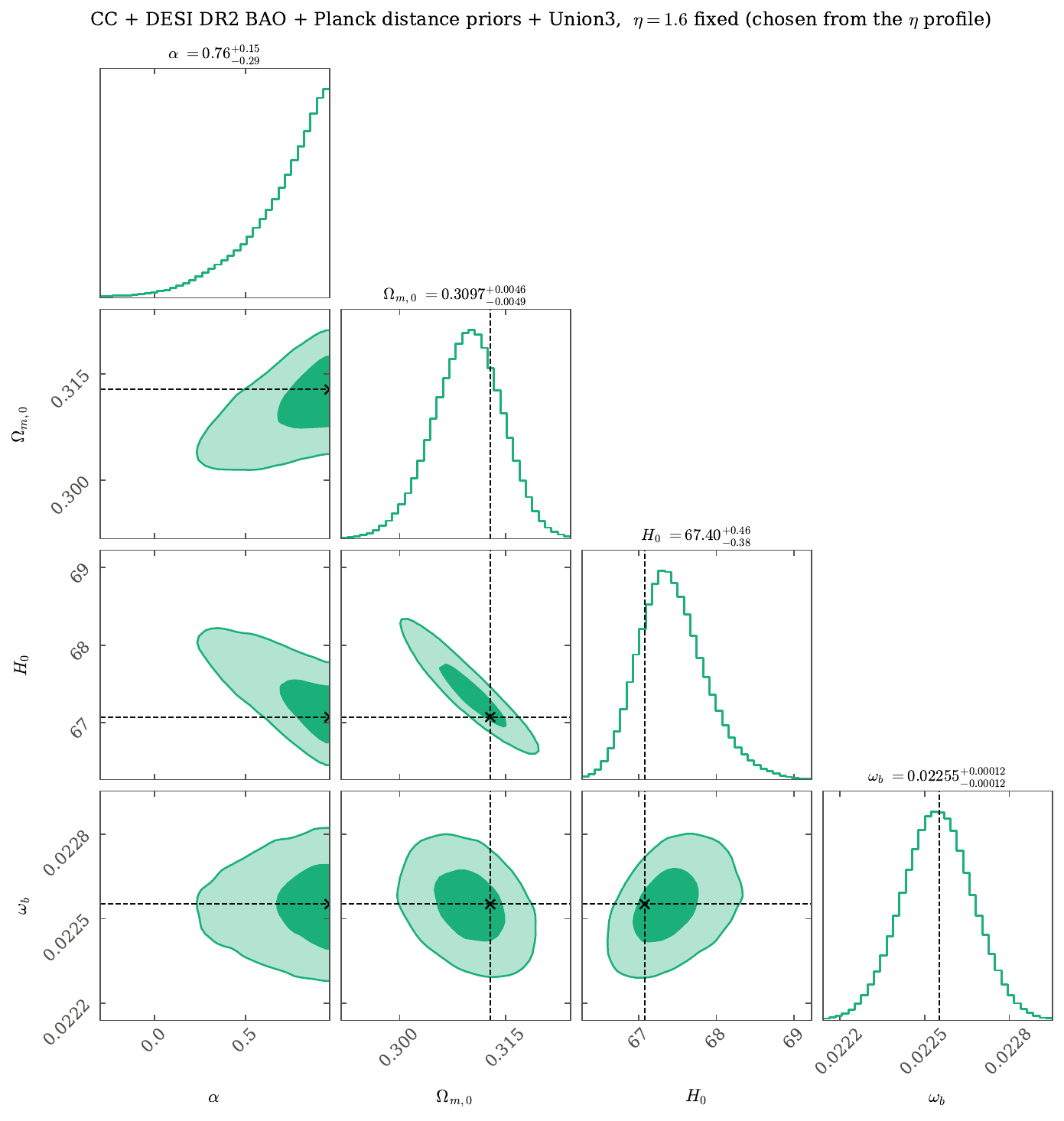}
\caption{As Fig.~\ref{fig:corner_du}, at $\eta=1.6$: DES-Y5 (left) and Union3 (right).}
\label{fig:corner_du16}
\end{figure*}

Figures~\ref{fig:corner_p16} and \ref{fig:corner_du16} show the posteriors at the data-driven benchmark $\eta=1.6$
for the three SN compilations. The posterior of $\alpha$ is broad and cut by the regularity bound $\alpha<0.962$, while
the other parameters are almost unchanged with respect to $\eta=\sqrt6$.

\section{Robustness to the analysis choices}
\label{app:robust}

Table~\ref{tab:robust} compares the improvement over \LCDM\ obtained with the analysis choices of this paper with those
obtained with an alternative set of choices (12 DESI points with QSO as $D_V$ only, the 1580 Pantheon+ SNe without the
Cepheid calibrators and $D_L=(1+z_{\rm HD})D_M$), with a fixed $z_*=1089.80$ in the CMB block, and with the growth data
of Sec.~\ref{sec:growth} added. The alternative choices change $\dchi$ by at most $0.45$ and the best-fit $\alpha$ by
at most $0.04$; the treatment of $z_*$ changes $\dchi$ by less than $0.05$, and the growth data lower it by $0.06$--$0.3$.

\begin{table}[h]
\centering
\caption{Improvement over \LCDM\ and best-fit $\alpha$ with the analysis choices of this paper (``this work''), with
the alternative choices described in the text (``alt.''), with a fixed $z_*$, and with the growth data added
($+f\sigma_8$, computed with $z_*$ at each step).}
\label{tab:robust}
\begin{adjustbox}{max width=\linewidth}
\begin{tabular}{llcccccc}
\toprule
SN & $\eta$ & $\dchi$ & $\dchi$ (alt.) & $\dchi$ (fixed $z_*$) & $\dchi$ ($+f\sigma_8$) & $\alpha_{\rm bf}$ & $\alpha_{\rm bf}$ (alt.) \\
\midrule
Pantheon+ & $\sqrt6$ & $2.02$ & $2.05$ & $2.04$ & $1.96$ & $0.087$ & $0.087$ \\
DES-Y5 & $\sqrt6$ & $2.86$ & $2.62$ & $2.89$ & $2.80$ & $0.095$ & $0.090$ \\
Union3 & $\sqrt6$ & $2.40$ & $2.45$ & $2.44$ & $2.29$ & $0.125$ & $0.124$ \\
Pantheon+ & $1.6$ & $4.79$ & $4.86$ & $4.76$ & $4.57$ & $0.835$ & $0.835$ \\
DES-Y5 & $1.6$ & $6.43$ & $5.99$ & $6.39$ & $6.18$ & $0.923$ & $0.883$ \\
Union3 & $1.6$ & $6.94$ & $7.09$ & $6.92$ & $6.65$ & $0.962$ & $0.962$ \\
\bottomrule
\end{tabular}
\end{adjustbox}
\end{table}

\section{Validation of the CMB block}
\label{app:cmb}

\citet{Chen2019} computed $R$ and $\ell_A$ for each sample of the Planck chains with the $z_*$ of that sample, so the
theory has to be evaluated at the $z_*$ of the sampled parameters. Table~\ref{tab:cmb} compares, at the best fits of
the main analysis, the decoupling redshift of CAMB~\citep{Lewis2000CAMB} with the fit used here, and gives $\ell_A$ and
the $\chi^2$ of the distance priors with a fixed and with the computed $z_*$. On a grid of 147 cosmologies with
$0.0218<\omega_b<0.0230$, $0.135<\omega_m<0.150$ and $66<H_0<70$ the fit reproduces the CAMB value of $z_*$ with an rms
error of $0.04$ (maximum $0.07$), and the formula of \citet{Aubourg2015}
reproduces $r_{\rm drag}$ to $0.03\%$. Since \citet{Chen2019} define $r_s(z_*)$ through an
integral that treats all matter as dust, we use the same integral rather than the sound horizon of CAMB, which differs from
it by $0.06\%$.

\begin{table}[h]
\centering
\caption{Planck distance priors at the best fits of the main analysis: $z_*$ from CAMB and from the fit of
Sec.~\ref{sec:data}, $\ell_A$ and $\chi^2_{\rm CMB}$ with a fixed $z_*=1089.80$ and with the computed $z_*$, at the same
parameters.}
\label{tab:cmb}
\small
\begin{adjustbox}{max width=\linewidth}
\begin{tabular}{lcccccc}
\toprule
Best fit & $z_*$ (CAMB) & $z_*$ (fit) & $\ell_A$ (fixed) & $\ell_A$ (fit) & $\chi^2_{\rm CMB}$ (fixed) & $\chi^2_{\rm CMB}$ (fit) \\
\midrule
Pantheon+, \LCDM & $1089.51$ & $1089.50$ & $301.441$ & $301.384$ & $3.68$ & $3.39$ \\
Pantheon+, $\eta=\sqrt6$ & $1089.40$ & $1089.38$ & $301.445$ & $301.368$ & $5.93$ & $5.22$ \\
Pantheon+, $\eta=1.6$ & $1089.44$ & $1089.43$ & $301.446$ & $301.377$ & $4.90$ & $4.34$ \\
DES-Y5, \LCDM & $1089.52$ & $1089.51$ & $301.439$ & $301.386$ & $3.41$ & $3.18$ \\
DES-Y5, $\eta=\sqrt6$ & $1089.39$ & $1089.38$ & $301.446$ & $301.367$ & $6.06$ & $5.32$ \\
DES-Y5, $\eta=1.6$ & $1089.44$ & $1089.42$ & $301.447$ & $301.377$ & $5.03$ & $4.43$ \\
Union3, \LCDM & $1089.50$ & $1089.49$ & $301.441$ & $301.384$ & $3.78$ & $3.47$ \\
Union3, $\eta=\sqrt6$ & $1089.37$ & $1089.35$ & $301.446$ & $301.363$ & $6.57$ & $5.74$ \\
Union3, $\eta=1.6$ & $1089.47$ & $1089.45$ & $301.445$ & $301.380$ & $4.47$ & $4.00$ \\
\bottomrule
\end{tabular}
\end{adjustbox}
\end{table}

\section{Monte Carlo calibration of the significance}
\label{app:lee}

Table~\ref{tab:lee} gives the result of the procedure of Sec.~\ref{sec:lee} for the three SN compilations, from
$2\times10^4$ mock data sets each. At fixed $\eta=\sqrt6$ the Monte Carlo and the $\chi^2$ significances coincide.
For the maximum over $(\alpha,\eta)$ the quantiles of the simulated distribution are $0.93$ (median), $4.8$
($95.4\%$) and $7.4$ ($99\%$), against $1.39$, $6.18$ and $9.21$ for a $\chi^2$ distribution with two degrees of
freedom and $0.45$, $4.00$ and $6.63$ for one: the search over $\eta$ multiplies the one-parameter $p$-value by a trials
factor of about $1.6$.

\begin{table*}
\centering
\caption{Significance of the improvement over \LCDM\ from $2\times10^4$ mock data sets generated under \LCDM. Left:
$\eta=\sqrt6$ fixed a priori, compared with a $\chi^2$ distribution with one degree of freedom. Right: maximum of $\dchi$
over $\alpha$ and $\eta$ (value of $\eta$ in brackets), compared with a $\chi^2$ distribution with two degrees of freedom.}
\label{tab:lee}
\small
\begin{adjustbox}{max width=\linewidth}
\begin{tabular}{lcccccccc}
\toprule
 & \multicolumn{4}{c}{$\eta=\sqrt6$} & \multicolumn{4}{c}{maximum over $(\alpha,\eta)$} \\
SN sample & $\dchi$ & $p_{\rm MC}$ & $n_\sigma^{\rm MC}$ & $n_\sigma^{(1)}$ & $\dchi_{\max}$ & $p_{\rm MC}$ & $n_\sigma^{\rm MC}$ & $n_\sigma^{(2)}$ \\
\midrule
Pantheon+ & $2.02$ & $0.157$ & $1.42$ & $1.42$ & $5.09$ ($1.5$) & $0.039$ & $2.06$ & $1.76$ \\
DES-Y5 & $2.86$ & $0.089$ & $1.70$ & $1.69$ & $6.49$ ($1.5$) & $0.017$ & $2.39$ & $2.06$ \\
Union3 & $2.40$ & $0.126$ & $1.53$ & $1.55$ & $7.17$ ($1.7$) & $0.012$ & $2.52$ & $2.20$ \\
\bottomrule
\end{tabular}
\end{adjustbox}
\end{table*}

\clearpage
\bibliographystyle{elsarticle-harv}
\bibliography{refs}

\end{document}